\documentclass[12pt,preprint]{aastex}

\begin{document}

\title{\emph{Spitzer} IRS Spectroscopy of \emph{IRAS}-Discovered Debris 
Disks$^{1}$}
\author{C.\ H.\ Chen\altaffilmark{2,3}, B.\ A. Sargent\altaffilmark{4},
        C.\ Bohac\altaffilmark{4}, K.\ H.\ Kim\altaffilmark{4},
	E.\ Leibensperger\altaffilmark{5}, M.\ Jura\altaffilmark{6}, 
	J.\ Najita\altaffilmark{2}, W.\ J.\ Forrest\altaffilmark{4}, 
	D.\ M.\ Watson\altaffilmark{4}, G.\ C.\ Sloan\altaffilmark{7}, 
	L.\ D. Keller\altaffilmark{5}}

\altaffiltext{1}{Based on observations with the NASA \emph{Spitzer Space
  Telescope}, which is operated by the California Institute of Technology
  for NASA}
\altaffiltext{2}{NOAO, 950 North Cherry Avenue, Tucson, AZ 85719; 
  cchen@noao.edu}
\altaffiltext{3}{Spitzer Fellow}
\altaffiltext{4}{Department of Physics and Astronomy, University 
  of Rochester, Rochester, NY 14627}
\altaffiltext{5}{Department of Physics, Ithaca College, Ithaca, 
  NY 14850}
\altaffiltext{6}{Department of Physics and Astronomy, University 
  of California, Los Angeles, CA 90095-1562}
\altaffiltext{7}{Center for Radiophysics and Space Research, 
  Cornell University, Ithaca, NY 14853-6801}

\begin{abstract}
We have obtained \emph{Spitzer Space Telescope} IRS 5.5 - 35 $\mu$m spectra
of 59 main sequence stars that possess \emph{IRAS} 60 $\mu$m excess. The 
spectra of five objects possess spectral features that are well-modeled using
micron-sized grains and silicates with crystalline mass fractions 0\% - 80\%,
consistent with T-Tauri and Herbig AeBe stars. With the exception of $\eta$ 
Crv, these objects are young with ages $\leq$50 Myr. Our fits require the 
presence of a cool black body continuum, $T_{gr}$ = 80 - 200 K, in addition to 
hot, amorphous and crystalline silicates, $T_{gr}$ = 290 - 600 K, suggesting 
that multiple parent body belts are present in some debris disks, analogous to 
the asteroid and Kuiper belts in our solar system. The spectra for the majority
of objects are featureless, suggesting that the emitting grains probably have 
radii $a$ $>$ 10 $\mu$m. We have modeled the excess continua using a continuous
disk with a uniform surface density distribution, expected if 
Poynting-Robertson and stellar wind drag are the dominant grain removal 
processes, and using a single temperature black body, expected if the dust is 
located in a narrow ring around the star. The IRS spectra of many objects are 
better modeled with a single temperature black body, suggesting that the disks 
possess inner holes. The distribution of grain temperatures, based on our black
body fits, peaks at $T_{gr}$ = 110 - 120 K. Since the timescale for ice 
sublimation of micron-sized grains with $T_{gr}$ $>$ 110 K is a fraction of a 
Myr, the lack of warmer material may be explained if the grains are icy. If 
planets dynamically clear the central portions of debris disks, then the 
frequency of planets around other stars is probably high. We estimate that the 
majority of debris disk systems possess parent body masses, $M_{PB}$ $<$ 1 
$M_{\earth}$. The low inferred parent body masses suggest that planet
formation is an efficient process. 
\end{abstract}


\section{Introduction}
Giant planets are believed to form in circumstellar disks by either (1) growth 
of interstellar grains into planetary cores and subsequent accretion of 
gas into planetary atmospheres \citep{pollack96} or (2) direct collapse
via gravitational instabilities. In the core-accretion model, the first step 
toward building planets is the growth of small bodies from sub-micron 
interstellar grains into meter-sized bodies. Fits of the 10 $\mu$m silicate 
features observed toward pre-main sequence T-Tauri and Herbig AeBe stars 
suggest that the grains in these systems have grown to radii of a few $\mu$m at
ages $<$few Myr (Forrest et al. 2004; Uchida et al. 2004; Honda et al. 2003; 
Bouwman et al. 2003). However, how and when micron-sized grains grow to 
kilometer-sized planetesimals, and the efficiency of this process are not well 
constrained. Mid-infrared spectroscopy of debris disks around main sequence 
stars may be used to detect larger grains and to infer the presence of 
kilometer-sized planetesimals at later ages.

Obtaining high signal:noise ground-based mid-infrared spectroscopy of debris 
disks has been challenging because typical debris disks have 10 $\mu$m 
fluxes, F$_{\nu}$ $<$ 1 Jy, making the majority of systems too faint to be 
studied spectroscopically from the ground and using \emph{ISO}. The excellent 
sensitivity of the IRS (Houck et al. 2004) on the \emph{Spitzer Space 
Telescope} (Werner et al. 2004) has enabled the spectroscopic study of large 
samples of debris disks, including all objects discovered with the \emph{IRAS} 
and \emph{ISO} satellites. We present the first 5 - 40 $\mu$m spectroscopic 
study of a large sample of debris disks to study the growth of large bodies in 
circumstellar disks and to elucidate the physical processes acting on 
micron-sized grains.

Debris disks are dusty, gas-poor disks around main sequence stars 
\citep{bp93, lba00, z01}. Micron-sized dust grains are inferred to exist in 
these disks from measurements of their thermal emission at infrared through 
millimeter wavelengths. The estimated lifetimes for circumstellar dust grains 
due to radiation and corpuscular stellar wind pressure (if the grains are 
small), sublimation (if the grains are icy), Poynting-Robertson and corpuscular
stellar wind drag, and collisions are typically significantly smaller than the 
estimated ages for the stellar systems, suggesting that the grains are 
replenished from a reservoir, such as sublimation of comets or collisions 
between parent bodies. 

The dust grains observed around debris disks may be produced when planets 
gravitationally perturb parent bodies producing collisions. In our solar 
system, the zodiacal dust possesses dust bands that thermally emit
$L_{IR}/L_{*}$ = 10$^{-7}$ times the incident stellar light and that have 
orbital properties identical to asteroid families, suggesting that the dust 
bands are generated by collisions between asteroids in each family. 
Gravitational perturbations by Jupiter and other planets in our solar system 
are expected to cause the apsides and nodes of asteroid orbits to precess at 
different rates because of small differences in their orbital parameters. This 
precession leads to asteroid collisions that generate the small grains observed
in the dust bands. However, observed debris disks are typically 3 - 5 orders of
magnitude more luminous than our zodical disk. Simulations of self-stirred 
disks suggest that the formation of icy planets, with radii 1000 - 3000 km, may
trigger collisional cascades between the remaining nearby kilometer-sized 
planetesimals \citep{kb04}. Giant planet migration, soon after formation, may 
also trigger collisions. The migration of the Jovian planets in our solar 
system during the Late Heavy Bombardment may have caused gravitational 
resonances to sweep through the main asteroid belt, sending asteroids into the 
inner solar system, producing the craters observed on old terrestrial planet 
surfaces \citep{str05}.

The presence of planets in debris disks may be inferred from the dynamical
influence they exert on dust particles. Planets may produce central clearings 
in disks by gravitationally scattering dust grains out of the system, that
are otherwise spiraling toward their orbit center under Poynting-Robertson and 
corpuscular stellar wind drag and by trapping grains into mean motion 
resonances \citep{lz99, qt02}. Unfortunately, directly resolving all but the 
nearest debris disks at infrared and submillimeter wavelengths, to search for 
disk structure generated by planets, is challenging with current ground- and 
space-based telescopes. However, the radial distribution of dust in a debris 
disk may be inferred from modeling the infrared spectral energy distribution 
(SED), assuming that the system is azimuthally symmetric.  

Mid-infrared spectra of 7 nearby debris disks, obtained with \emph{Spitzer} 
IRS, have revealed that the majority of these systems do not possess spectral
features, suggesting that the grains probably have $a$ $>$ 10 $\mu$m 
(Jura et al. 2004; Sloan et al. 2004). Jura et al. (2004) and Sloan et al. 
(2004) model the excess continuua of these objects using a continuous disk with
a uniform surface density, expected if Poynting-Robertson drag is the dominant 
dust removal mechanism, and using a single temperature grain model. The disks 
around two-thirds of the stars in their sample appear to possess inner 
truncations at 10 - 50 AU. One possible explanation for the presence central 
clearings is a planet at the truncation distance that sweeps the inner regions 
of the disk clear; however, central clearings may also be produced by 
sublimation if the grains are icy or by radiation pressure if the disks are
collisionally dominated and grains with sub-blow out sizes are removed by
radiation pressure. The SED's of the remaining systems are consistent with a 
disk with a uniform surface density that may be produced when large grains, 
generated by collisions between parent bodies, spiral into their orbit center 
under the Poynting-Robertson effect (and stellar wind drag).

We report the results of a \emph{Spitzer} IRS study of 59 main sequence stars 
with published \emph{IRAS} 60 $\mu$m excesses, building on initial results 
published by Jura et al. (2004). Jura et al. (2004) included LL spectra 
obtained for 19 debris disks. This study includes SL and LL or SH/LH 
data, depending on the \emph{IRAS} 25 $\mu$m flux, and expands the target 
sample size. Refinements in the IRS flats have allowed us to improve the 
signal:noise at which the spectra are measured; therefore, we reanalyze objects
in Jura et al. (2004) in addition to presenting new data. We list the targets
for the full sample, along with their spectral types, distances, and published 
ages in Table 1.

\section{Sample Selection and Characterization}
The debris disks around Vega, Fomalhaut, $\epsilon$ Eridani, and $\beta$
Pictoris were initially discovered from the presence of strong \emph{IRAS}
60 $\mu$m and 100 $\mu$m excesses, 10 - 100 times larger than expected from the
photosphere alone (Backman \& Paresce 1993). Studies comparing the \emph{IRAS} 
fluxes with predictions for the photospheric emission of field stars 
subsequently discovered more than 100 debris disk candidates, corresponding to 
a disk fraction $\sim$15\% \citep{bp93,cot87,mb98,sn86, syl96,ww88}.

We selected 115 stars with \emph{IRAS} 60 $\mu$m excesses, discovered in the 
studies listed above, for further study with the IRS. Approximately 25 of these
targets are pre-main sequence Herbig Ae/Be stars that will be discussed in 
L. Keller et al. (2006, in preparation) and another 30 of these targets are
extended in the IRS slit, suggesting that the excess emission is generated by 
interstellar grains, and will be described in B. Sargent et al. (2006, in 
preparation). The discovery of extended infrared excess around two objects 
(HR 1307, HR 2522) is consistent with coronagraphic imaging that revealed the 
presence of reflection nebulosities in these systems\citep{kal02}. Modeling of 
the scattered light images suggest that the dust grains are small, interstellar
grains, located at distances 1,000 - 100,000 AU, rather than large 
circumstellar grains at distances $<$100 AU from the central star 
\citep{kal02}.

We list the debris disks in our study with their stellar properties and 
reported \emph{IRAS} 25, 60, and 100 $\mu$m excesses in Table 1. The majority 
of the stars in our sample are nearby (within 150 pc), isolated main sequence 
stars; although a couple of objects may lie outside the local bubble and 
another nine objects are members of well-studied OB Associations or moving 
groups. For example, the star $\lambda$ Cas is a member of Cas-Tau 
\citep{dez99} with an estimated age of 10 Myr \citep{bha00}. The star 
HD 146897 is a member of Upper Scorpius and the stars HD 95086, G Cen, 
HD 110058, and HD 113766 are members of Lower Centaurus Crux in Sco-Cen 
(de Zeeuw et al. 1999) with estimated ages of 5 Myr and 16 Myr \citep{mml02}, 
respectively. The stars HR 7012, $\eta$ Tel, and HD 181327 are members of the 
$\beta$ Pic moving group with estimated ages of $\sim$12 Myr \citep{zsb01}. 

Since we would like to study the evolution of dust properties as a function of 
time, we estimate ages for as many stars in the sample as possible by fitting 
$\log g$ and $T_{eff}$ to \cite{sch92} isochrones for B-, A-type, and F-type 
stars. We estimate $\log g$ and $T_{eff}$ (see Table 1) from mean General 
Catalogue of Photometic Data Stromgren photometry \citep{mmh97} using the 
calibration of \cite{nap93} and the rotation correction of \cite{fb98}. B-, A-,
and F-type stars which appear above the published isochrones are assigned ages 
of 1 Myr, 50 Myr, and 300 Myr, respectively, the ages for which the main 
sequence isochrones begin. We estimate uncertainties in our estimated ages of a
factor of two. For comparison, we also list published isochronal and moving 
group ages in Table 1. For cases in which the moving group age is inconsistent 
with the isochronal age, we assume that the moving group ages are more accurate
because stars above the zero age main sequence in the HR diagram may be either 
evolving down onto the main sequence or up and away from the main sequence.

\section{Observations}
We obtained IRS spectra of 59 main-sequence stars with previously reported IRAS
60 $\mu$m excesses \citep{bp93,cot87,mb98,sn86, syl96,ww88} with either 
(1) both the Short-Low (5.2 - 14.0 $\mu$m) and Long-Low (14.0 - 38.0 $\mu$m; 
$\lambda$/$\Delta \lambda$ $\sim$ 90) modules or (2) the Short-Low, Short-High 
(9.9 - 19.6 $\mu$m), and Long-High (18.7 - 37.2 $\mu$m; 
$\lambda$/$\Delta \lambda$ $\sim$ 600) modules. In order to avoid 
time-consuming peak-up on our relatively bright targets with accurately known 
positions, we operated the observatory in IRS spectral mapping mode where a 
2 $\times$ 3 raster (spatial $\times$ dispersion) centered on the star is 
performed (Watson et al. 2004). We carried out the bulk of the reduction and 
analysis of our spectra with the IRS team's SMART program (Higdon et al. 2004).

We estimate the stellar photospheric fluxes of our objects by minimum
$\chi^{2}$ fitting published photometry from the literature to model stellar
atmospheres, using only bandpasses with wavelengths shorter than 3 $\mu$m:
TD 1 \citep{tho78}, \cite{joh66}, and 2MASS (Cutri et al. 2003). For stars with
spectral types earlier than K2V, we use 1993 Kurucz stellar atmospheres; for 
stars later than K2V, we use Nextgen models. We assume that all stars have 
solar abundances and $\log g$ = 4.5 unless otherwise noted. We use the 
\cite{ccm89} extinction law to estimate $E(B-V)$ and list the extinctions 
($A_{V}$ = 3.1 $E(B-V)$) estimated from photosphere fitting in Table 1. We plot
the SEDs for all of our objects in Figure 1. Ultra-violet through near-infrared
photometry are shown with black symbols; IRS spectra are shown in red; MIPS 
photometry, where available (Bryden et al. 2005; Rieke et al. 2005), are shown 
with blue error bars; IRAS photometry are shown with green error bars and 
upper limit symbols; and submillimeter photometry, where available (Wyatt
et al. 2005, Greaves et al. 2004, Sheret et al. 2003, Holmes et al. 2003), are 
shown with black error bars and upper limit symbols. Our photosphere models are
shown with a solid black line. 

We measure the flux of our objects in two photometric bands to search for 
excess emission from silicates (8.5-13 $\mu$m) and cold grains (30-34 $\mu$m). 
The calibration uncertainty in the fluxes is $\sim$5\% and the measured 
statistical uncertainties are listed in Table 2. We find 25 sources without 
strong IRS excesses despite reported \emph{IRAS} 60 $\mu$m excesses (annotated 
in Table 2). The average fractional excess, 
($F_{\nu}$(measured)-$F_{\nu}$(predicted))/$F_{\nu}$(predicted) for these 
photospheric objects is slightly negative in the 8.5 - 13 $\mu$m band
(-0.028$\pm$0.055) with a standard deviation consistent with IRS observations 
of nearby, solar-like stars (Beichman et al. 2006). The average fractional 
excess is slightly higher in the 30 - 34 $\mu$m band and possesses a larger 
standard deviation (0.08$\pm$0.13), indicating that some of these sources may 
possess weak excesses at the longest IRS wavelengths. For sources that are
not saturated and do not possess excess in the 8.5 - 13 $\mu$m band, we also 
normalized the photosphere to the first 10 data points in the SL module to 
search more sensitively for excesses in the 30 - 34 $\mu$m band. Values of this
calibration factor deviated from unity on a star-by-star basis by less than 
5\%. Pinning the photosphere to the fluxes at the shortest wavelengths of the 
SL module produces average fractional excesses for the 25 stars without IRS 
excesses of -0.032$\pm$0.048 and 0.06$\pm$0.13 at 8.5-13 and 30-34 $\mu$m, 
respectively. In our analysis, we use photosphere models that are scaled to the
first ten points of SL unless the source is saturated or possesses a 8.5 - 13 
$\mu$m excess.

The significantly larger \emph{IRAS} beam may contain multiple sources and
therefore be responsible for some of the discprepancy between the \emph{IRAS}
and \emph{Spitzer} IRS results. For reference, the \emph{IRAS} beam is 
$1\arcmin \times 5 \arcmin$ at 12 $\mu$m and 25 $\mu$m, 
$2\arcmin \times 5\arcmin$ at 60 $\mu$m, and $4\arcmin \times 5\arcmin$ at 
100 $\mu$m. By contrast, point sources observed with \emph{Spitzer} MIPS at 
have a FWHM of $\sim$6$\arcsec$ at 24 $\mu$m and a FWHM of $\sim$20$\arcsec$
at 70 $\mu$m. The color-corrected \emph{IRAS} 25 $\mu$m fluxes are 
significantly higher than the IRS 25 $\mu$m fluxes for three objects (HR 2124, 
HR 6297, and HD 200800). HR 6297 has been observed with MIPS at 24 $\mu$m. In 
this case, the discrepancy between the HR 6297 \emph{IRAS} and IRS fluxes can 
definitely be explained by the presence of a second source in the \emph{IRAS} 
beam. The \emph{IRAS} Point Source Catalog 25 $\mu$m flux for HR 6297 is (not 
color-corrected) $\sim$410 mJy (with a central position 0.4$\arcmin$ away from 
the star) while the IRS 25 $\mu$m flux is $\sim$90 mJy. MIPS 24 $\mu$m images 
reveal two sources near HR 6297; one at the position of HR 6297 with 
$F_{\nu}$(23.68 $\mu$m) = 82 mJy, consistent with the IRS spectrum, and 
another, brighter source 0.92$\arcmin$ west of HR 6297 with $F_{\nu}$(23.68 
$\mu$m) = 207 mJy. The \emph{IRAS} PSC flux is the sum of the two MIPS sources 
and is therefore an overestimate of the HR 6297 25 $\mu$m flux. Unfortunately, 
the remaining two sources with possibly confused \emph{IRAS} 25 $\mu$m fluxes, 
have not been observed with MIPS. We searched the 2MASS catalog around the 
positions of HR 2124 and HD 200800 for additional sources of confusion. For 
comparison, we examined the 2MASS colors for the unexpected 25 $\mu$m source 
(located at 17:00:00.07 -54:35:38.6 (J2000)) in the HR 6297 \emph{IRAS} beam 
(the spurious source possesses $J$=6.69, $H$= 5.73, and $K$=5.38). We found a 
bright source with similarly red 2MASS colors, 0.9$\arcmin$ away from HD 200800
(located at 21:09:13.83 -65:47:38.4 (J2000)) with $J$=6.40, $H$= 5.47, and 
$K$=5.17. However, we were not able to find an obvious candidate for the extra 
emission in the \emph{IRAS} HR 2124 beam. There are 7 2MASS sources within 
2$\arcmin$ of HR 2124 with $J-K$ $>$ 1 but all have $K$ $\sim$ 14-15, 
suggesting that they are probably too faint to be detected at 24 $\mu$m despite
their red colors.

The presence of \emph{IRAS} 60 $\mu$m excess does not necessarily imply the 
presence of 5 - 35 $\mu$m excess. Twenty-five objects in our sample 
($\sim$42\%) do not possess strong infrared excess at 5 - 35 $\mu$m despite
the identification of \emph{IRAS} excess in the literature. The discrepancy 
between published \emph{IRAS} results and our results may (1) reflect 
difficulties in either the search criteria or the data reduction in the 
original searches, (2) be due to the presence of cold dust that is not detected
at IRS wavelengths, or (3) be due to source confusion in the \emph{IRAS} beam. 
Eleven of the objects in our study possess \emph{IRAS} 12 $\mu$m and 25 $\mu$m 
fluxes that are consistent with our photosphere models and \emph{IRAS} upper 
limits or photosphere detections in the \emph{IRAS} Point Source or Faint 
Source Catalogs at $\lambda$ $\geq$ 60 $\mu$m (107 Psc, CC Eri, $\tau^{1}$ Eri,
$\alpha$ For, HR 1686, $\delta$ Dor, HR 3220, FI Vir, HR 3862, G Cen, and 
$\mu$ Ara) despite reports of measured excess in the literature. The 
discprepancy in these cases may be due to difficulties in the search criteria
or data reduction. Our sample is drawn from a several \emph{IRAS} searches for 
debris disks that used different criteria to establish whether a source 
possessed an infrared excess. Cot\'{e} (1986) selected sources with $V$-[12],
$V$-[25],$V$-[60] $>$0.5; Sadakane \& Nishida (1988) selected sources with 
[12]-[60] $>$ 1; and Walker \& Wolstencroft (1988) selected sources with 
$F_{\nu}$(60 $\mu$m)/$F_{\nu}$(100 $\mu$m) = 0.8-2.0, similar to that observed 
toward $\beta$ Pic, Vega, Fomalhaut, and $\epsilon$ Eridani. 

The dust in many of the undetected systems may be cold. The IRS spectra for 
five objects are photospheric over the bulk of the IRS wavelength range but may
be rising at $\lambda$ $\sim$35 $\mu$m although with insufficient signal:noise 
to model these systems in detail ($\kappa$ Lep, HR 2124, XZ Lep, $\gamma$ Tra, 
and HD 200800). The dust in some systems may be so cold that excess emission 
can not be easily detected above the photosphere at IRS wavelengths. Ten 
sources possess \emph{IRAS} 60 $\mu$m flux excesses but no evidence for 
significant IRS excess at $\lambda$ $<$ 35 $\mu$m ($\tau$ Cet, $\gamma$ Dor, 
32 Ori, $\psi^{5}$ Aur, $\delta$ Vel, HR 5236, $\sigma$ Boo, 41 Ser, 
$\sigma$ Her, HD 221354). If the dust in these systems has $T_{dust}$ $<$ 60 K,
then the infrared excess at the longest IRS wavelengths could be small and 
changing slowly compared to the stellar photosphere, making the excess 
difficult to infer. For $\tau$ Cet, our measured and predicted fluxes, 
$F_{\nu}$(8.5-13 $\mu$m) = 7470 and 7710 mJy, respectively, and 
$F_{\nu}$(30-34 $\mu$m) = 786 and 829 mJy, respectively, suggesting that this 
source possesses -3\% and -5\% excesses at 8.5-13 and 30-34 $\mu$m, 
respectively. However, the 60 $\mu$m - 850 $\mu$m excesses observed toward 
$\tau$ Cet are easier to detect because the photosphere is negligible at
these wavelengths; these excesses are fit by a modified 60 K black body 
\citep{gre04}, implying that the slope of the 20 $\mu$m - 35 $\mu$m spectrum 
should be $\sim$4\% higher than without the excess. Alternately, the 
\emph{IRAS} flux for some objects may include extragalactic or other stellar 
sources that were included in the large \emph{IRAS} beam.

\section{Grain Composition}
Five objects in our sample possess 10 $\mu$m and/or 20 $\mu$m spectral 
features: HR 3927, $\eta$ Crv, HD 113766, HR 7012, and $\eta$ Tel. We model the
excess emission for these objects assuming that the observed features are 
generated by amorphous olivine and pyroxene, and crystalline forsterite, 
enstatite, and silica. We model the remaining continuum emission using 
amorphous carbon and one (or two) black body distributions. We do not 
observe PAH emission features toward any of the objects in our sample.

A lower limit to the size of dust grains orbiting a star can be found by 
balancing the force due to radiation pressure with the force due to gravity. 
For small grains with radius $a$, the force due to radiation pressure 
overcomes gravity for solid particles larger than
\begin{equation}
a_{min,o} = 3 L_{*} Q_{pr}/(16 \pi G M_{*} c \rho_{s})
\end{equation}
\citep{art88} where $L_{*}$ and $M_{*}$ are the stellar luminosity and 
mass, $Q_{pr}$ is the radiation pressure coupling coefficient, and $\rho_{s}$
is the density of an individual grain. We assume that the bulk density of 
silicate, amorphous carbon, and silica grains are $\rho_{s}$ = 3.3, 2.5, and 
2.3 g cm$^{-3}$, respectively. Since radiation from A-type (HR 3927, HR 7012, 
and $\eta$ Tel) and F-type ($\eta$ Crv and HD 113766) stars is dominated by 
optical and ultraviolet light, we expect that $2 \pi a/\lambda \gg 1$ and 
therefore the effective cross section of the grains can be approximated by 
their geometric cross section so $Q_{pr} \approx 1$. We estimate the stellar 
mass by fitting our inferred $T_{eff}$ and $\log g$ to the \cite{sch92} 
isochrones for A-type stars and by fitting our inferred $T_{eff}$ and $L_{*}$ 
to \cite{sdf00} isochrones for F-type stars. We list the minimum grain sizes, 
$a_{min,o}$, for solid silicate, amorphous carbon, and silica grains around 
each object if the species is required in the fit in Table 3. If the grains are
porous (with a volume fraction, $f=V_{vac}/V_{tot}$ $>$ 0), then the 
minimum-sized grains estimated in equation (1) must be modified to account for 
the vacuum fraction:
\begin{equation}
a_{min}(f) = \frac{a_{min,o}}{1-f}
\end{equation}
In fitting the 5 - 40 $\mu$m spectra of HR 3927, $\eta$ Crv, HD 113766, 
HR 7012, and $\eta$ Tel, we assume that the grains are spheres with radius, 
$a$ $>$ $a_{min}$. We list the solid angles subtended by each dust population 
$\Omega$, their mass $m$ and temperature, the radii of each grain population, 
and its vacuum volume fraction in Table 3. We infer absorption coefficients, 
$Q_{abs}(\lambda)$, using optical constants published in the literature for 
amorphous olivine (MgFeSiO$_{4}$, Dorschner et al. 1995), amorphous pyroxene 
(Mg$_{0.5}$Fe$_{0.5}$SiO$_{3}$ and Mg$_{0.8}$Fe$_{0.2}$SiO$_{3}$, Dorschner et 
al. 1995), crystalline forsterite (Mg$_{1.9}$Fe$_{0.1}$SiO$_{4}$, Fabian et al.
2001), crystalline enstatite (MgSiO$_{3}$, Jaeger et al. 1998), and amorphous 
carbon \citep{zub96}, and Bruggeman Effective Medium Theory \citep{bh83}. 
(Please see Sargent et al. 2006 for a more detailed description of 
$Q_{abs}(\lambda)$ estimates.) Non-zero vacuum volume fractions shift the
peak positions of silicate features to longer wavelengths and broaden the 
features (Kessler-Silacci et al. 2006).

Our fits to the HD 113766 and HR 7012 spectra require the presence of small 
sub-$\mu$m-sized grains (crystalline enstatite, silica, and amorphous carbon).
In these cases, we modeled the sub-micron dust components using laboratory 
measured opacities of crushed forsterite (Mg$_{2}$SiO$_{4}$, Koike et al. 2003)
and enstatite (Mg$_{0.7}$Fe$_{0.3}$SiO$_{4}$, Chihara et al. 2002) and optical 
constants for silica (cristobalite, Simon \& McMahon 1953) assuming that the 
distribution of grain shapes is well-described by the continuous distribution 
of ellipsoids (CDE; Fabian et al. 2001). Both of these systems are young with 
estimated ages $\sim$16 and $\sim$12 Myr, respectively. The disks around 
around both stars are optically thin and possess high fractional infrared
luminosities, $L_{IR}/L_{*}$ = 0.015 and 10$^{-3}$, respectively. Since the
estimate ages of these systems are $\sim$16 and $\sim$12 Myr, based on their 
membership in Lower Centaurus Crux in Sco-Cen and the $\beta$ Pictoris Moving 
Group, these systems probably no longer contain bulk molecular gas. Therefore, 
we hypothesize that the small particles in each system may have been generated 
in a recent collisions between parent bodies. 

We are confident in the identification of silicate species around HD 113766, 
HR 7012, and $\eta$ Crv, however, the exact dust mass in amorphous olivine,
pyroxene and carbon is somewhat uncertain. For example, for HD 113766, the 
presence of sharply peaked spectral features at 10.0, 11.1, 12.0, 16, 19, and 
23.5 $\mu$m are used to identify forsterite. For HR 7012, the presence of 
spectral features at 9.3 and 10.5 $\mu$m are used to identify crystalline 
pyroxene. Once we fit the sharply peaked crystalline features, then we add 
amorphous olivine and pyroxene to fit the overall silicate feature; finally,
we add amorphous carbon and one (or two) black bodies to fit the remaining 
continuum. The fit to the non-crystalline component of the spectrum is 
degenerate and may be fit with different ratios of amorphous silicates. In 
addition, the emissivity of amorphous carbon is approximately constant at 5 
- 35 $\mu$m, suggesting that a black body could be substituted for this 
component. In the case of HR 7012, we use cristobalite because it produces a 
better fit to the 8.7 $\mu$m shoulder of the 10 $\mu$m silicate feature than 
alpha quartz; however, other silicates such as opal (hydrous silicate) may also
provide good fits.

The majority of debris disks with spectral features possess crystalline 
silicate emission features; only $\eta$ Tel in our sample does not. We estimate
crystalline silicate mass fractions of 76\% and 0\% for HR 7012 and $\eta$ Tel 
(two $\sim$12 Myr old members of the $\beta$ Pic moving group), 4.1\% for 
HD 113766 (a $\sim$16 Myr old member of Sco-Cen), 38\% for HR 3927 (a 
$\sim$50 Myr old field A-type main sequence star), and 31\% for $\eta$ Crv (a 
$\sim$1 Gyr old field F-type main sequence star). The crystalline silicate 
fraction for HR 7012 appears extremely high; however, those of the other debris
disks are consistent with measurements toward pre-main sequence T-Tauri and 
Herbig AeBe stars (Sargent et al. 2006; Bouwman et al. 2001). Therefore, we 
find no correlation between age and crystallinity for circumstellar 
silicates. The most dramatic example of the lack of correlation between 
crystalline silicate fraction and age is the disparity in the crystallinity of 
the dust around HR 7012 and $\eta$ Tel, both A-type main sequence members of 
the $\beta$ Pic moving group \citep{zsb01}.

Our solar system is believed to possess two populations of dust grains produced
by collisions in two distinctive small body belts: (1) the zodiacal dust 
located at $D$ = 2 - 4 AU with $T_{gr}$ = 270 K (Reach et al. 2003), generated 
by collisions between asteroids and (2) another population of cooler dust at 
$D$ = 30 - 50 AU, with $T_{gr}$ = 50 - 60 K, that is believed to be generated 
by collisions between objects in the Kuiper Belt; this population of dust has 
not been detected directly thus far. Each of our systems with spectral features
possesses at least two distinctive dust populations: a hot population with 
$T_{gr}$ = 290 - 600 K that is inferred to exist from detailed fitting of 
amorphous and crystalline silicate features, and a cooler population with 
$T_{gr}$ = 80 - 200 K that is inferred to exist from black body fits to the 
excess continuum. In addition, $\eta$ Crv possesses a third, even colder dust 
component with $T_{gr}$ = 40$\pm$5 K, inferred from SED fits to \emph{IRAS} 
12 - 100 $\mu$m and JCMT SCUBA 450 and 850 $\mu$m photometry \citep{wyaetal05}.
Our results establish the presence of multiple belts of small bodies in 
debris disks; previously, the evidence was much weaker. The ``hot'' dust 
component in these systems is only 2.5 - 3.6$\times$ hotter than the ``cold'' 
dust component, while zodiacal dust is estimated to be 4 - 5$\times$ hotter 
than dust in the Kuiper Belt. HR 3927, $\eta$ Crv, HD 113766, HR 7012, 
and $\eta$ Tel may possess asteroid and Kuiper belts in analogy with our solar 
system at larger distances from their central stars because of their higher 
luminosity. For $\eta$ Crv, the detection of three planetesimal belts with 
black body distances of 1.3, 11, and 102 AU respectively may imply the
presence of planets at 2.6, 22, and 204 AU (comparable to the distances of 
Mars, Saturn, and Sedna) if these belts are stirred by planets in the same way 
that the main asteroid belt is stirred by Jupiter in our solar system. 
Detailed dynamical models of this system are required to determine whether 
$\eta$ Crv possesses multiple planets.

\section{Dust Properties}
The majority of infrared excess sources in our sample apparently lack spectral 
features, suggesting that the grains are too cold or too large ($a$ $>$ 
10 $\mu$m) to produce features. We fit, using $\chi^2$ minimization, the 
5 - 35 $\mu$m photosphere subtracted spectra for stars without silicate 
emission features, assuming that the grains are black bodies. The photosphere 
subtracted IRS spectra for all excess sources without spectral features are 
shown in Figure 3. The majority of our objects were observed in spectral 
mapping mode in which the slit is moved across the source, producing six 
independent spectra for each object. The error bars in this figure represent 
the difference between the two spectra in which the source is best centered. 
We overlaid our best fit single temperature black body model in blue. We list 
the best fitting grain temperatures, $T_{gr}$, and fractional infrared 
luminosities, $L_{IR}/L_{*}$, inferred for each system assuming $L_{IR}$ = 
4$\Omega \sigma T_{gr}^{4} d^{2}$, where $\Omega$ is the solid angle subtended 
by the grains (in steradian) and $d$ is the distance from the Sun to the 
central star, in Table 4 with the reduced $\chi^{2}$ for each fit. Since no 
significant excess is detected toward $\tau$ Cet, despite a strong 
submillimeter excess characterized by $T_{gr}$ = 60 K, populations of cold dust
may not be detected using IRS. Similarly, the temperatures inferred for cool 
dust populations using IRS spectra may be inaccurate.

The minimum grain distance can be constrained from the grain temperature, 
$T_{gr}$, assuming that the dust particles act like black bodies. Black bodies 
in radiative equilibrium with a stellar source are located a distance
\begin{equation}
D = \frac{1}{2} (\frac{T_{eff}}{T_{gr}})^{2} R_{*}
\end{equation}
from the central star \citep{jur98}, where $T_{eff}$ and $R_{*}$ are the 
effective temperature of the stellar photosphere and the stellar radius. We 
estimate the stellar temperatures and absolute V-band magnitudes from Stromgren
photometry using the calibration of \cite{nap93} corrected for rotation 
\citep{fb98} and infer luminosities using the bolometric correction from 
\cite{flo96}. We estimate $R_{*}$ assuming $L_{*}$ = 
$4 \pi R_{*}^{2} \sigma T_{*}^{4}$. The black body grain distances (listed in 
Table 4) range between 4 AU for the K1V star HD 53143 and 72 AU for the B8Vn 
star $\lambda$ Cas.

We estimate the average grain size of orbiting dust grains assuming that 
radiation pressure removes grains with radii $< a_{min,o}$ (please see equation
1) and that the dust grain size distribution is determined by collisional 
equilibrium \citep{gn89}
\begin{equation}
n(a) da = n_{o} a^{-p} da
\end{equation}
with p $\simeq$ 3.5, similar to that inferred for the interstellar medium
based on interstellar extinction curves, even though the grain-size 
distributions in these systems are not well constrained. Interplanetary and
lunar size distributions with $p_{small}$ = 2.2 - 3.7 for grains with
$a$ = 0.3 - 2 $\mu$m and $p_{large}$ = 2.0 for grains with
$a$ = 2 - 20 $\mu$m have been used to reproduce \emph{ISO} 5 - 16 $\mu$m
observations  of zodiacal dust (Reach et al. 2003). If we weight by the 
number of particles, we estimate average grain sizes, $<a>$ = 5/3 $a_{min,o}$ 
(see Table 4), assuming that the density of an individual grain, $\rho_{s}$ 
(= 2.5 g cm$^{-3}$). The majority of the stars in our sample are B-, A-, and 
F-type stars. (HD 53134 is the only object with infrared excess discussed here 
with spectral type later than F.)  Since the radiation from these objects is 
dominated by optical and ultraviolet light, we expect that $2 \pi a/\lambda \gg
1$ and therefore the effective cross section of the grains can be approximated 
by their geometric cross section so $Q_{pr} \approx 1$. We estimate the stellar
mass by fitting our inferred $T_{eff}$ and $\log g$ to the \cite{sch92} 
isochrones for B- and A-type stars and by fitting our inferred $T_{eff}$ and 
$L_{*}$ to \cite{sdf00} isochrones for the remaining stars. We find average 
grain sizes between 0.2 $\mu$m for HD 53143 and 25 $\mu$m, for the B8Vn star 
$\lambda$ Cas (see Table 4), similar to those inferred for the zodiacal dust in
our solar system (Reach et al. 2003).

We can estimate the minimum mass of dust around objects in our sample assuming 
that the particles have radius $a_{min,o}$; if the grains are larger, then our 
estimate is a lower bound. If we assume a thin shell of dust at distance, $D$, 
from the star and if the particles are spheres of radius, $a$, and if the 
absorption cross section of the particles equals their geometric cross section,
then the mass of dust is
\begin{equation}
M_{dust} \geq \frac{16}{3} \pi \frac{L_{IR}}{L_{*}} \rho_{s} D^{2} a_{min,o}
\end{equation}
\citep{jur95} where $L_{IR}$ is the luminosity of the dust. We estimate 
dust masses in micron-sized infrared emitting grains between 
3.7$\times$10$^{-8}$ $M_{\earth}$ for HD 53143 and 2.3$\times$10$^{-4}$ 
$M_{\earth}$ for the A3IV/V star HR 1082 (see  Table 4). If the grains
possess a size distribution $n_{o} a^{-3.5}$ with a maximum radius
$a_{max}$ = 10 cm, as inferred from submillimeter observations (Zuckerman et 
al. 1995), we can estimate a the dust mass in larger grains. If the measured 
excess flux is $F_{\nu}(excess)$ and the black body flux for the excess is 
$B_{\nu}(excess)$, at frequency, $\nu$, then
\begin{equation}
M_{10 cm} = \frac{4}{3} \rho \sqrt{a_{min,o} a_{max}} \ d^{2} \
\frac{F_{\nu}(excess)}{B_{\nu}(excess)}
\end{equation}
where $d$ is the distance to our sun. We use the solid angle subtended by the 
dust grains, inferred from the minimum $\chi^{2}$ black body fits to the 
excess, $\Omega$, to determine the flux to black body ratio, 
$\frac{F_{\nu}(excess)}{B_{\nu}(excess)}$ = $\Omega$ and list the estimated
larger grain dust masses in Table 4. 

\section{Dust Removal Mechanisms}
Dust grains in debris disks may be removed by a variety of processes such
as radiation and corpuscular stellar wind pressure, ice sublimation, and 
collisions. Collisions may shatter parent bodies into small grains that are 
radiatively drive grains from the system. Larger grains in these high density 
environments may continue to collide until they reach sizes below the blow-out 
limit and are radiatively ejected. In lower density disks around B- and A-type
main sequence stars, large grains may be subject to the Poynting-Robertson 
effect and may spiral in toward the central star. In the absence of planets 
in our solar system, Poynting-Robertson and corpuscular solar wind drag would 
determine the spatial distribution of dust. The discovery of debris disks 
around lower-mass solar-like and M-type stars has led to speculation that 
corpuscular stellar winds may contribute to grain removal in a manner analogous
to radiation pressure and the Poynting-Robertson effect \citep{pjl05}: (1) An 
outflowing corpuscular stellar wind produces a pressure on dust grains which 
overcomes the force due to gravity for small grains. (2) Particles orbiting the
star are subject to a drag force produced when dust grains collide with atoms 
in the stellar wind. These collisions decrease the velocity of orbiting dust 
grains and therefore their angular momentum, causing them to spiral into the 
central star.

The dominant grain removal process within a disk is dependent not only on 
the luminosity of the central star but also on grain distance from the central 
star. For Fomalhaut, \cite{bp93} estimate that at 67 AU collisions to sizes 
below the blow-out limit (grains below the blow-out limit are quickly 
radiatively ejected from the system) are the most effective grain removal
mechanism while at 1000 AU the Poynting-Robertson effect is the most efficient 
grain removal process. In Figure 4, we plot the sublimation lifetime, the 
Poynting-Robertson (and corpuscular stellar wind) drag lifetime, and the 
collision lifetime for average-sized grains around typical B5V, A5V, and F5V 
stars. Sublimation may quickly remove icy grains in the innermost portions of 
the disk. At larger radii, collisions dominate grain destruction, and at the 
largest radii, where the disk has the lowest density, Poynting-Robertson and 
corpuscular stellar wind drag may dominate grain destruction. We estimate the 
lifetimes of average-sized grains and the parent body masses around stars in 
our sample if ice sublimation, Poynting-Roberton and corpuscular stellar wind 
drag, and collisions are each the dominant grain removal mechanism in the 
absence of other processes.

\subsection{Ice Sublimation}
One possibile explanation for the presence of central clearings, inferred
from black body fits to the IRS spectra, is that the grains are icy and
sublimate when they come too close to the central star. Although the dust grain
composition can not be determined directly from spectral features, it may be 
inferred from the statistical grain properties in our sample. We plot the 
distribution of inferred grain temperatures in Figure 5. The estimated grain 
temperatures appear to cluster between 110 K and 130 K. Laboratory studies
find that thermal desorption of water ice (H$_{2}$O) from H$_{2}$O layers 
begins at temperatures of 120 K and is completed by 170 K \citep{fcmw01}.
Therefore, the peak in grain temperature at 110 K to 120 K may suggest that
the grains are icy and are beginning to sublimate. 

If the grains are icy, then sublimation may also remove grains from the 
disk. If the grain temperature is constant while the star remains on the
main sequence, then we may write the following expression for the sublimation 
lifetime of an average grain. 
\begin{equation}
t_{subl} = \frac{<a> \rho_{i} T_{gr}^{1/2} e^{T_{subl}/T_{gr}}}
{\dot{\sigma_{o}}}
\end{equation}
\citep{jur98} where $\dot{\sigma_{o}}$ is the mass rate per surface area 
(= 3.8 $\times$ 10$^{8}$ g cm$^{-2}$ s$^{-1}$ K$^{1/2}$, $T_{subl}$ = 5530 K; 
Ford \& Neufeld 2001) and $\rho_{i}$ = 1.5 g cm$^{-3}$ if the grains are mostly
ice with some refractory material mixed in, as expected for Kuiper Belt 
objetcs. The average grain radius, $<a>$, listed in Table 4 is computed 
assuming that the grains are composed of silicates with $\rho_{s}$ = 2.5 g 
cm$^{-3}$; therefore, if the grains are mainly composed of ices, then the
estimated average grain radii are an underestimate and the values in this table
should be multiplied by 5/3. We list the sublimation lifetimes for grains in 
Table 5 assuming that they are icy. The sublimation lifetimes are sensitively 
dependent on grain temperature. For example, 3.5 $\mu$m grains around HR 1082 
with an estimated $T_{gr}$ = 70 K, have a sublimation lifetime, $T_{subl}$ = 
1.3 $\times$ 10$^{7}$ Gyr while 16 $\mu$m grains around HR 6211 with an 
estimated $T_{gr}$ = 160 K, have a sublimation lifetime, $T_{subl}$ = 7.4 
minutes. In systems with $T_{gr}$ $>$ 100 K, the sublimation lifetime is the 
shortest lifetime by more than an order of magnitude. 

\subsection{Poynting-Robertson Drag}
If Poynting-Robertson (PR) drag is the dominant grain removal mechanism, then 
grains spiral in from the radii at which they are created toward their orbit 
center, creating a continuous disk with uniform surface density and a $1/D$ 
volume density. In this model, the inner radius of the disk coincides with the 
stellar radius unless the grains sublimate, or are dynamically ejected by a 
massive body interior to the radius at which the grains are produced. For an 
optically thin, gas-free disk whose particle density, $n$, varies as $D^{-q}$, 
the infrared spectrum should be well-described by the function
\begin{equation}
F_{\nu} = K_{1} B_{\nu}(T_{*}) + K_{2} \nu^{-3 + 2q + 0.5pq - 0.5p}
\end{equation}
\citep{jur98} where the absorption coefficient for the grains, $Q_{abs}$ 
$\propto$ $\nu^{p}$ and $K_{1}$ and $K_{2}$ are constants. The first term 
describes the photospheric emission and the second the infrared excess. If the 
grains are large (p = 0) and the surface density is determined by PR drag 
(q = 1), then the infrared spectrum, $F_{\nu}$ $\propto$ $\nu^{-1}$. We plot 
the minimum $\chi^{2}$ fits to the IRS photosphere subtracted spectra in blue 
in Figure 3, assuming that the grains are large ($p$ = 0) and list the fitting 
parameter $K_{2}$ in Table 4 along with the minimum reduced $\chi^{2}$ for the 
fits. The continuous disk model has a lower reduced $\chi^{2}$ than the single 
temperature black body for one object in our sample (HR 8799); however, this 
source possess a weak IRS excesses ($<$0.1 Jy) which is detected with a SNR 
$<$ 5.

Our simple SED models suggest that PR drag may not be the dominant grain 
removal mechanism in debris disks. To test this hypothesis, we compare the PR 
drag lifetimes and the lifetimes for grains under sublimation, corpulscular 
stellar wind drag, and collisions for all of the sources in our study. The PR 
drag lifetime of grains in a circular orbit, a distance $D$ from a star is
\begin{equation}
t_{PR} = \left( \frac{4 \pi <a> \rho_{s}}{3} \right) \frac{c^{2} D^{2}}{L_{*}}
\end{equation}
\citep{bur79}. The PR drag lifetimes of average-sized grains (12000 yr - 
1.1 Myr; see Table 5), estimated using the grain properties in Table 4, are 
significantly shorter than the stellar ages ($t_{age}$), suggesting that the 
grains are replenished through collisions between larger bodies. The PR drag
lifetime of average-sized grains around HR 8799, the only object whose 
excess spectrum is better modeled by F$_{\nu}$ $\propto$ $\lambda$, is not the 
shortest grain lifetime by a factor of 2; therefore, we do not expect that
this system should possess a uniform disk. We estimate lower limits for the 
parent body masses around our objects assuming that all of the grains are 
destroyed by PR drag and that the systems are in steady state. If $M_{PB}$ 
denotes the mass in parent bodies, then we may write
\begin{equation}
M_{PB} \geq \frac{4 L_{IR} t_{age}}{c^2}
\end{equation}
\citep{cj01}. We estimate the infrared luminosities of the systems from the 
black body fits to the excess. If the grains emit a substantial fraction of 
their radiation at $\lambda$ $>$30 $\mu$m, then this approximation may not
be valid. The inferred parent body masses range between 1.3 $\times$ 10$^{-3}$ 
$M_{\earth}$ for the K1V star HD 53143 and 0.93 $M_{\earth}$ for the A3IV/V 
star HR 1082 (see Table 5). 

Whether PR drag or other processes, such as collisions, are the dominant grain 
destruction mechanism in debris disks is uncertain and depends on the density 
of dust grains and the spatial distribution of dust for each particular object.
Numerical models of the dynamical evolution of dust in collisional equilibrium
suggest that Poynting-Robertson drag is the primary mechanism for grain
transport in disks with $L_{IR}/L_{*}$ $\leq$ 10$^{-6}$ while radiation
pressure is the primary mechanism for dust transport in collisionally
dominated disks with $L_{IR}/L_{*}$ $\geq$ 10$^{-4}$ (Krivov et al. 2000).
Recently, \cite{wya05} concluded that PR drag is relatively unimportant in 
\emph{IRAS}-discovered debris disks. He found that the volume density of grains
in \emph{IRAS}-discovered debris disks is so high that grains which migrate 
inward under Poynting-Robertson drag will suffer destructive collisions with 
other grains and will be rapidly expelled by radiation pressure. The 
destruction of inward-migrating dust grains via mutual collisions may 
explain the presence of central clearings without requiring the presence
of icy grains or planets in debris disks.

\subsection{Stellar Wind Drag}
For B-type main sequence stars and young solar-like main sequence stars, both
of which possess strong stellar winds, drag on dust grains produced by loss of 
angular momentum to corpuscular stellar wind may be stronger than that produced
by the Poynting-Robertson effect. Stellar wind drag may explain the observed 
anti-correlation between \emph{Spitzer} 24 $\mu$m excess and \emph{ROSAT} 
fluxes toward F-type stars in the 3 - 20 Myr Sco-Cen \citep{chen05} and
the lack of 12 $\mu$m excesses observed toward nearby, $>$10 Myr-old, late-type
M-dwarfs \citep{pjl05}. Recently, \cite{sc06} have reproduced the radial 
brightness profile of the AU Mic disk assuming that collisions between parent 
bodies on circular orbits at 43 AU produce the observed dust grains. Large 
grains produce a surface density, $\sigma$ $\propto$ $r^{0}$, at r $<$ 43 AU, 
under corpuscular and Poynting-Robertson (CPR) drag modified by collisions 
while small grains that are barely bound under corpuscular stellar wind and 
radiation pressure produce a surface density, $\sigma$ $\propto$ $r^{-5/2}$, in
the outer disk. 

Stellar winds around B-type stars are produced by the transfer of momentum
from photons below the photosphere to material outflowing in the wind. The 
mass loss rates and expansion velocities can be measured from radio 
observations of free-free emission produced in the outer parts of the wind
or from H$\alpha$ emission which may be formed in the inner regions of the
wind. Monte Carlo simulations of stellar winds that include multiple 
scattering of photons, successfully reproduce stellar wind mass loss rates 
measured at radio wavelengths \citep{vin00}. We estimate the stellar mass loss 
rates (${\dot M_{wind}}$) for the B-type stars in our sample using a fit to the
numerical models of \cite{vin00} that depend on the stellar luminosity, mass, 
and effective temperature, assuming that the ratio of the terminal wind 
velocity to the effective wind escape velocity at the stellar surface, 
$v_{\infty}/v_{esc}$ = 1.3. The increase in ``drag'' in the inward drift 
velocity of dust grains under corpuscular stellar wind and Poynting-Robertson 
drag over that produced by Poynting-Robertson drag alone is given approximately
by the factor (1 + ${\dot M_{wind}} c^{2}/L_{*}$) \citep{jura04}. 

The mass loss rates due to stellar winds around 14 nearby, solar-like stars 
have been inferred via Lyman $\alpha$ absorption, produced when the stellar 
wind collides with the surrounding interstellar medium, producing a hot 
\ion{H}{1} astrosphere with an effective temperature 20,000 - 40,000 K 
\citep{woo02, woo05}. \cite{woo02, woo05} fit the stellar mass loss rate, 
${\dot M_{wind}}$, per stellar surface area, $A$, as a power-law function of 
x-ray flux per stellar area, ${\dot M_{wind}}/A \propto F_{x}^{1.34\pm0.18}$, 
assuming that the wind speed for solar-like stars is similar to that measured 
for the Sun, $v_{wind}$ = 400 km s$^{-1}$. The uncertainty in the 
${\dot M_{wind}}$ extrapolation is probably a factor of two because the size of
the astrosphere and the amount of astrospheric absorption scales as the square 
root of the wind ram pressure; the wind ram pressure, $P_{wind}$ $\propto$ 
$\dot{M}_{wind} v_{wind}$; and the variation in the solar wind speed is 
approximately a factor of two. The stellar mass loss rate power law dependence 
on x-ray flux per stellar area saturates at $F_{x}$ = 8$\times$10$^{5}$ ergs 
cm$^{-2}$ s$^{-1}$. One possible explanation for the saturation is that 
stars possess more polar spots as they become more magnetically active,
indicating changes in the field geometry. Changes in the field structure
could precipitate changes in the stellar wind \citep{woo05}.

We infer ${\dot M_{wind}}$ from \emph{ROSAT} fluxes for our sample assuming the
\cite{woo05} power-law, scaling to observations of $\alpha$ Cen ($F_{x}$ 
= 3.7 $\times$ 10$^{4}$ erg cm$^{-2}$ s$^{-1}$, ${\dot M_{wind}}/A$ = 
0.9 ${\dot M_{\sun}}/A_{\sun}$); the astrosphere for this source is 
well-detected in Lyman $\alpha$ and its ${\dot M_{wind}}/A$ and $F_{x}$ lie on 
the published fit. We list the observed \emph{ROSAT} fluxes, HR 1 hardness 
ratios between the 0.1 - 0.4 and the 0.5 - 2.0 keV bands, and the angular 
offsets between the \emph{ROSAT} catalog sources and the FGKM-type stars in our
sample in Table 6. The x-ray spectra for stars in our sample have flat spectral
energy distributions, with a mean hardness ratio -0.4, consistent with 
observations of stars in the solar neighborhood. We estimate X-ray 
luminosities, $L_{X}$, using the conversion 1 \emph{ROSAT} count = 
(8.31 + 5.30 HR1) $\times$ 10$^{-12}$ ergs cm$^{-2}$ \citep{fsg95}. The
inferred $F_{x}$ $>$ 8$\times$10$^{5}$ ergs cm$^{-2}$ s$^{-1}$ for 8 solar-like
and low-mass stars, suggesting that the stellar winds for these stars are 
probably one or two orders of magnitude smaller than inferred from the 
\cite{woo05} relation: $\tau^{1}$ Eri, $\alpha$ For, HD 53143, HD 113766, 
HD 139664, HD 146897, HD 181327, and HD 191089. We do not make any 
extrapolations for the mass loss rates in these systems.

For the B-type stars and solar-like and low mass stars with $F_{x}$ $<$ 
8$\times$10$^{5}$ ergs cm$^{-2}$ s$^{-1}$, ${\dot M_{wind}} c^{2}/L_{*}$ $>$ 1,
suggesting that stellar wind drag can not be neglected (see Table 5). The 
lifetime for grains in a circular orbit under the Poynting-Robertson effect and
stellar wind drag is
\begin{equation}
t_{PR+wind} = \frac{1}{\left( \frac{\dot{M_{wind}} c^2}{L_{*}}+1 \right)}  
t_{PR} 
\end{equation}
We estimate the combined Poynting-Robertson and stellar wind drag lifetimes
using the Poynting-Robertson drag time lifetimes and 
${\dot M_{wind}} c^{2}/L_{*}$ values listed in Table 5. For B-type stars,
${\dot M_{wind}} c^{2}/L_{*}$ is typically $\sim$2; therefore, the drag 
lifetimes of the grains are reduced in most cases by $\sim$60\%. For example,
$\lambda$ Cas and HR 6532 possess ${\dot M_{wind}} c^{2}/L_{*}$ = 1.4 and 1.2,
respectively. However, for F-type and later stars, 
${\dot M_{wind}} c^{2}/L_{*}$ is typically $\sim$100 if the source is detected 
by \emph{ROSAT}; therefore, the drag lifetimes of the grains are reduced by a 
factor of a couple hundred. For example, HR 506 and HR 6670 may possess 
${\dot M_{wind}} c^{2}/L_{*}$ as high as 82 and 190. Unfortunately, 
\emph{ROSAT} upper limits do not place stringent upper limits on 
${\dot M_{wind}} c^{2}/L_{*}$ in cases in which the source is not detected.
The mass in parent bodies assuming that the system is in steady state and 
that Poynting-Robertson drag and stellar wind drag are the dominant grain 
removal mechanisms is given by the expression,
\begin{equation}
M_{PB} \geq \left( \frac{\dot{M_{wind}} c^2}{L_{*}} + 1 \right) 
\frac{L_{IR}}{c^2} t_{age}
\end{equation}
We estimate the mass in parent bodies using the inferred stellar properties
listed in Table 4. For B-type stars, the stellar wind drag inferred parent body
masses are consistent with the PR drag inferred parent body masses, $M_{PB}$ 
$<$1 $M_{\earth}$. However, for solar-like stars, the stellar wind drag 
inferred parent body masses are substantially (20 - 50$\times$) higher but
still less than 1 $M_{\earth}$.  The fact that so little material remains in
parent bodies suggests that planet formation must be an efficient process
if planets have already formed in these systems. 

\subsection{Collisions}
If the particle density within the disk is high, then collisions are expected
to dominate grain destruction by generating small grains that are removed 
rapidly via radiation and corpuscular stellar wind pressure. We estimate
the collision lifetime assuming that grains are on inclined orbits, such that 
they encounter the surface density of the disk twice per orbit, that tangential
collisions are destructive, and that they have a collisional equilibrium size 
distribution (given in equation 4)
\begin{equation}
t_{coll} = 3000 \ \textup{yr} (D/30 AU)^{7/2} 
(M_{submm}/0.1 M_{\earth})^{-1} (\sqrt{a_{min} a_{max}}/1 \ \textup{mm})
(M_{\sun}/M_{*})^{1/2}
\end{equation}
\cite{bp93} using grain distances, dust masses (extrapolated for micron-sized 
through centimeter-sized grains), and minimum radii in Table 4. For the 
majority of our systems, the collision lifetimes computed are comparable to or 
shorter than the Poynting-Robertson and/or stellar wind drag lifetimes (see 
Table 5), suggesting that collisions dominate the destruction of particles, 
consistent with published studies \citep{wya05, dd03, nw05}. In Figure 4, we 
plot the sublimation lifetime if the grains are icy, the Poynting-Robertson and
stellar wind drag lifetime, and the collision lifetime as a function of radius 
around typical B5V, A5V, and F5V stars. In all cases, sublimation is the 
dominant grain removal process at small distances from the star. For typical 
A5V and F5V stars, the collision lifetime for average-sized grains is shorter 
than the drag lifetime if the disk has a dust mass between 0.001 $M_{\earth}$ 
and 1 $M_{\earth}$, even if the F5V star has a stellar wind with a mass loss 
rate as high as $\dot{M}_{wind}$ = 1000 $\dot{M}_{\sun}$. However, for a 
typical B5V star, the Poynting-Robertson and stellar wind drag lifetime may be 
shorter than the collision lifetime, especially at large radii, if the disk has
a dust mass, $M_{submm}$ $\sim$ 0.1 $M_{\earth}$; however, we infer small 
$M_{submm}$ = 1.0$\times$10$^{-5}$ $M_{\earth}$ for some stars in our sample, 
suggesting that Poynting-Robertson and corpuscular stellar wind drag effects 
should be the dominant grain removal mechanism for at least some objects.


The fractional infrared luminosity of a debris disk for a fixed distance is 
expected to decrease inversely with time, $L_{IR}/L_{*}$ $\propto$ 1/$t_{age}$,
if collisions are the dominant grain removal process and is expected to 
decrease inversely with time squared, $L_{IR}/L_{*}$ $\propto$ 1/$t_{age}^{2}$,
if Poynting-Robertson drag is the dominant grain removal process \citep{dd03}. 
We plot $L_{IR}/L_{*}$ as a function of age for the systems in our sample in 
Figure 6a. For young stars which do not appear on the \cite{sch92} isochrones, 
we assume moving group ages rather than our assigned ages. The upper envelope 
of the distribution can be fitted with the function, 
$L_{IR}/L_{*}$ = $(L_{IR}/L_{*})_{o}$ ($t_{o}/t_{age}$), where 
$(L_{IR}/L_{*})_{o} t_{o}$ = 0.40 Myr; however, the function $L_{IR}/L_{*}$ = 
$(L_{IR}/L_{*})_{o}$ ($t_{o}/t_{age}$)$^{2}$, with 
$(L_{IR}/L_{*})_{o} t_{o}^{2}$ = 60 Myr$^{2}$ does not produce a bad fit. 
The $1/t^{2}$ trend line includes all of the sources in our sample except
$\eta$ Crv which has a high $L_{IR}/L_{*}$ = 3$\times$10$^{-4}$ for its age of 
1 Gyr. Our data set is consistent with the idea that debris disks are 
collisionally dominated systems but our data set is too small to determine 
whether Poynting-Robertson drag dominates grain removal at any particular age.
A 1/$t_{age}$ time dependence with a decay timescale $t_{o}$ = 150 Myr 
has been observed for the MIPS 24 $\mu$m and \emph{IRAS} 25 $\mu$m excess 
around $\sim$270 A-type stars, consistent with grain destruction via collisions
(Rieke et al. 2005). Similarly, A 1/$t_{age}$ time dependence with a decay 
timescale $t_{o}$ = 200 Myr has been observed for the 850 $\mu$m excess
around 13 nearby, solar-like stars also consistent with grain destruction
via collisions \citep{nw05}. Our study differs from these in the Rieke et al. 
(2005) and the \cite{nw05} samples in that (1) our sample contains a 
heterogenous mix of stars with varying spectral type and (2) our data set 
includes the measurement of the excess at more than one wavelength; the 
measurement of the SED allows us to infer the dust temperature and therefore 
the infrared luminosities and dust masses/dust production rates more 
accurately.

In a minimum mass solar nebula, 1000 - 3000 km-sized bodies are expected
to grow on timescales, $t_{P}$ $\approx$ 15 - 20 Myr ($D$/30 AU)$^{3}$
\citep{kb04,kb05} and to perturb planetesimals in the disk, initiating 
collisional cascades that produce micron-sized grains. This model predicts a 
$t^{-0.6}$ to $t^{-0.35}$ decay of the fractional infrared luminosity for the 
whole debris disk at ages of 10 Myr to 1 Gyr, depending on the tensile strength
of the grains, a somewhat more shallow evolution of the fractional infrared 
luminosity than inferred by Dominik \& Decin (2003). The Kenyon \& Bromley 
(2004, 2005) models also predict that older debris disks systems possess 
infrared bright rings of dust at larger radii than their younger counterparts. 
Since the dust grains in our sample probably have $a$ $>$ 10 $\mu$m, we plot 
the inferred black body dust distance, $D$, as a function of stellar age to 
test this hypothesis (see Figure 6b). The black body distances, inferred from 
SED models, are consistent with the measured radii of HR 4796A, Fomalhaut, and 
Vega from maps of thermal emission from large particles at mid-infrared and 
submillimeter wavelengths to within a factor of 2 (Holland et al. 2004; Wilner 
et al. 2002; Jayawardhana et al. 1998). However, the black body distances are 
smaller than the measured radii of $\beta$ Pic and AU Mic in scattered light 
(Krist et al. 2005; Heap et al. 2000) and than the measured radius of Vega, 
inferred from maps of thermal emission of stochasitically heated small grains 
(Su et al. 2005). We do not find a clear correlation between dust grain 
distance and the age of the central star, in agreement with submillimeter 
studies \citep{nw05}. The addition of submillimeter data for $\eta$ Crv and 
$\tau$ Cet does not appear to improve our fit of grain distance as a function 
of stellar age. In Figure 6b, we overplot grain distances for these objects 
inferred from submillimeter SED models.

\section{Correlations Between Stellar and Dust Properties?}
We searched for correlations between stellar and dust properties. We 
investigated whether the multiplicity, the rotational velocity, or the
metallicity of the central star is correlated with either the observed
fractional infrared luminosity or grain temperature.

In a binary system, the orbits of dust grains at distances approximately 1.6
to 2.6 times the binary separation are expected to be unstable (Artymowicz \& 
Lubow 1994), leading to the formation of two populations of circumstellar dust 
grains. Disks around each component of the binary system are trucated at their 
outer radii but cool dust may reside at much larger distances in circumbinary 
disks. We searched for a correlation between the single/binary nature of a 
stellar system and the inferred circumstellar dust properties (grain 
temperature and fractional infrared luminosity) to determine whether the 
gravitational effects of a secondary star affect disk properties. 

Eleven stars in our sample are binary systems; five of which (45.5\%) 
possess IRS excesses detected with good signal-to-noise ratios (SNRs). One of
the binary systems with IRS excess does not possess spectral features and is 
better modeled using a single temperature black body than a uniform disk whose 
surface density is given by Poynting-Robertson Drag (HR 1570). The grain 
temperature for this systems is $T_{gr}$ = 90.6 K and the fractional infrared 
luminosity for this system is 3.0$\times$10$^{-5}$. For comparison, forty-nine 
stars in our sample are single systems; twenty-nine of which (59.2\%) possess 
IRS excesses detected with good SNRs. Fifteen of the single systems with IRS 
excesses do not possess spectral features and are better modeled using a single
temperature black body (HR 333, HR 506, $\gamma$ Tri, HR 1082, HD 53143, 
HR 3314, HD 95086, HD 110058, $\lambda$ Boo, HD 139664, HD 146897, HR 6297, 
HR 6486, HD 181327, and HD 191089). The mean grain temperature for these 
systems is $T_{gr}$ = 103 K with a standard deviation of 19 K. The mean 
fractional infrared luminosity of these systems is 8.4$\times$10$^{-4}$, with a
standard deviation of 1.5$\times$10$^{-3}$. The mean grain temperatures and 
fractional infrared luminosities for single and binary systems are consistent 
when the standard deviation of these quantities are taken into account. 

One difference appears when the single and binary star populations are 
compared. The fraction of infrared excess binary systems which can be modeled 
by a single temperature black body is significantly smaller than the number of 
infrared excess single star systems. We list the angular and physical 
separation of all the binary systems in our sample in Table 7. One possibility 
for the poor single temperature black body fits associated with binary systems 
is the presence of multiple populations of circumstellar dust grains (e.g., 
circumprimary, circumsecondary, and circumbinary disks). If the binary nature 
of these systems contributes to the complicated structure of the infrared 
excess then objects which are not well modeled by a single temperature black
body should possess binary separations 1 - 100 AU. However, the separations of
one of the two systems ($\lambda$ Cas, HR 6532) which are not well modeled by a
single temperature  black body is too small to truncate a circumstellar disk. 
HR 6532 is a spectroscopic binary with a period of 6.8 days. Larger statistics
are needed to confirm whether the infrared spectra of binary systems is
more complicated than a single temperature black body.

We plot the fractional infrared luminosity and grain temperature as a function 
of binary separation (see Figure 7) to examine further the effects of a 
secondary star. We assume very small separations for Algol eclipsing and 
spectroscopic binaries ($<$1 AU). Systems to the left of the dotted vertical 
line have separations $<$10$\arcsec$, small enough that both the primary and 
secondary fall into the LL (and sometimes the SL) slit. Systems to the right of
the dotted verticle line are too widely separated for the secondary to 
contribute to either the SL or LL spectrum. The average fractional infrared 
luminosity for systems with separations of 10 - 100 AU, 1.6$\times$10$^{-5}$, 
is smaller than that inferred for objects with separations of 100 - 500 AU, 
7.6$\times$10$^{-4}$; however, this comparison is highly biased by the very 
high fractional infrared luminosity associated with HD 113766, 
1.5$\times$10$^{-2}$, because our binary sample is so small. The average grain 
temperature for systems with separations of 10 - 100 AU, 120 K, is consistent
with the peak in grain temperature distribution for the whole sample, but
smaller than that inferred for systems with separations of 100 - 500 AU, 
240 K. Debris disks with $T_{gr}$ $\sim$ 250 K are rare \citep{ap91}. The very 
high average $T_{gr}$ estimated is based on a sample of two unusual objects 
(HD 113766 and $\eta$ Tel). Detailed modeling of the 10 $\mu$m feature observed
toward these objects suggests black body grains with $T_{gr}$ $>$ 300 K and 
amorphous olivine with $T_{gr}$ $>$ 200 K (see section 4; Schutz, Meeus, \& 
Sterzik 2005). 

X-ray studies of solar-like stars suggest that magnetic breaking causes stars 
to rotate more slowly as they age, with a $v_{rot}$ $\propto$ t$^{-0.6\pm0.1}$ 
time dependence for stars with ages $>$0.3 Gyr \citep{aye97}. We searched
for correlations between the inferred circumstellar dust properties and the 
measured projected stellar rotational velocity, $v \sin i$. Using $v \sin i$
as a proxy for stellar age, we expect that both the fractional infrared 
luminosity and the grain temperature will decrease with stellar age or increase
with $v \sin i$. We plot the fractional infrared luminosity and grain 
temperature as a function of $v \sin i$ in Figure 7, using different symbols 
for B-, A-, F-, and K-type stars. For binary systems, we used the $v \sin i$ 
for the primary star. Stars with spectral type earlier than F5V are shown with 
open symbols while stars with spectral type F5V and later are shown with solid 
symbols. Early-type stars are not expected to spin down as they age because 
they do not possess deep convective envelopes. When F5V and later spectral-type
stars in our sample are compared with one another, a possible decrease in 
fractional infrared luminosity as a function of $v\sin i$ and increase in
grain temperature as a function of $v\sin i$ are seen. However, our sample is 
predominantly B- and A-type stars and only contains 3 mid-F and later type 
stars with measured dust grain properties and measured $v\sin i$; therefore, no
strong conclusions about the dependence of dust grain properties on $v\sin i$ 
can be drawn in this study.

Radial velocity studies of main sequence stars that possess giant planets 
find a correlation between the presence of an orbiting planet and the 
metallicity of the central star. Spectral synthesis modeling of high resolution
visual spectra, sensitive to stellar semiamplitudes $>$ 30 m s$^{-1}$ and 
orbital periods shorter than 4 yr, find that fewer than 3\% of stars with 
-0.5 $<$ [Fe/H] $<$ 0.0 have Doppler-detected planets while 25\% of stars with 
[Fe/H] $>$ +0.3 possess giant planets \citep{fv05}. Recent \emph{Spitzer} MIPS 
observations suggest that nearby planet-bearing, solar-like stars may be more 
likely to possess 70 $\mu$m excesses and larger average 70 $\mu$m excesses than
stars without known planets (Beichman et al. 2005). These excesses are 
generated by cool dust ($T_{gr}$ $<$ 100 K) located beyond 10 AU, well outside 
the orbits of the discovered planets. If correlations exist between metallicity
and the presence of a planet and between the presence of a planet and a 
70 $\mu$m excess, there might also be a correlation between the presence of IRS
excess and stellar metallicity; although, no correlations between the presence 
of 70 $\mu$m excess and stellar metallicity have been found thus far (Bryden et
al. 2006; Beichman et al. 2005). 

We plot fractional infrared luminosity and grain temperature versus [Fe/H] for 
the eleven F-type stars in our sample that possess [Fe/H] measurements 
determined from the Stromgren photometry survey of Nordstrom et al. (2004; see 
Figure 7). For the six objects that do not possess IRS excess, we indicate 
their position along [Fe/H] axis with an upper limit symbol. The stars with and
without detected IRS excess have similar metallicities, consistent with a 
stochastic origin for the small infrared emitting grains in debris disks. 
We do not include B- and A-type stars in this plot because they possess shallow
surface convective zones. The observed metallicity of these objects may be more
easily distorted by pollution than their F-type counterparts. We observe a 
large dispersion in both the fractional infrared luminosity and grain 
temperature even though our sample only contains 5 F-type stars with measured 
grain properties and measured [Fe/H]. More objects from our sample may be 
folded in when grain properties inferred from longer wavelength observations 
are included; for example, $\tau$ Cet, with [Fe/H] = -0.47, 
log($L_{IR}/L_{*}$) = -4.6 \citep{dec03} and $T_{gr}$ = 60 K \citep{gre04}
inferred from far-infrared and submillimeter photometry. 

\section{Gas Mass Upper Limits}
Atomic and molecular gases may affect the dynamics of circumstellar dust 
grains. Numerical models of disks with gas:dust ratios of 0.1 - 10 suggest 
that gas-grain interactions may generate the observed infrared bright rings by
concentrating small grains, with radii just above the blowout radius, at
the outer edge of the gas disk \citep{ta01}. In addition, measurement of gas 
masses in a large sample of disks may help determine the gas dissipation 
timescale and help constrain models for giant planet formation
(Hollenbach et al. 2005). We searched for H$_{2}$ S(0), S(1) and \ion{S}{1} 
emission at 28.2 $\mu$m, 17.0 $\mu$m, and 25.2 $\mu$m, respectively, to 
constrain the bulk gas mass in debris disks. We do not detect any of these 
emission lines in any of our SH and LH module data. We estimate 3$\sigma$ upper
limits to the H$_{2}$ S(0), S(1) and \ion{S}{1} line fluxes from our nod 
difference spectra (see Table 8). Since the amplitude of each difference 
spectrum varied from pixel to pixel near each line, we averaged each difference
spectrum in a region $\pm$0.5 $\mu$m around each line to determine the 
uncertainty in the line flux. We estimate H$_{2}$ mass upper limits from the 
S(0) line flux upper limits, assuming that the source is unresolved and that 
the gas has a temperature, $T_{ex}$ = 50 K and 100 K, expected from bulk gas
that is co-spatial with the infrared-emitting dust. Since the most constraining
gas mass upper limits are typically $<$100 $M_{\earth}$ and the measured 
submillimeter dust masses are typically $<$1 $M_{\earth}$, we can not constrain
the gas:dust ratio in debris disks well.

Molecular gas has been detected toward the nearby (70 pc away from the Sun) 
$\sim$20 Myr old star 49 Cet. Submillimeter observations find CO 
J = 3$\rightarrow$2 and J = 2$\rightarrow$1 emission at the radial velocity of 
the star \citep{dgc05, zfk95}. Models of the 49 Cet doubly
peaked CO J = 3$\rightarrow$2 profile are consistent with a compact disk with 
an outer radius, $D_{out}$ = 17 AU, and a disk inclination, $i$ = 16$\arcdeg$; 
the high velocity wings of the profile ($\sim$4 km sec$^{-1}$ relative to the 
star) suggest that gas is present at $\leq$5 AU from the central star 
\citep{dgc05}. \emph{ISO} observations detect H$_{2}$ S(0) and S(1) emission 
with a line flux 6.6$\pm$2.0 $\times$ 10$^{-14}$ ergs sec$^{-1}$ cm$^{-2}$ and 
place 3$\sigma$ upper limits on the S(1) line flux, $<$0.8 $\times$ 10$^{-14}$ 
ergs sec$^{-1}$ cm$^{-2}$, corresponding to a gas:dust ratio $>$100 (Thi et al.
2001). Our 49 Cet 3$\sigma$ H$_{2}$ S(0) line flux upper limit apparently 
conflicts with the Thi et al. (2001) results because our S(0) upper limit is a 
factor $\sim$9 lower than the reported detection. Our upper limit is consistent
with the mass expected in $H_{2}$ inferred from 800 $\mu$m - 1100 $\mu$m 
continuum measurements of the dust mass, $M_{submm}$ = 0.02 - 1 $M_{\earth}$, 
if the disk has an interstellar gas:dust ratio (Zuckerman et al. 1995) and is 
also consistent with the inferred CO masses from submillimeter measurements. 
\cite{zfk95} and \cite{dgc05} estimate that between 2 and 30 $M_{\earth}$ 
H$_{2}$ exists in the disk, assuming a gas excitation temperature, 
$T_{ex}$ = 50 - 60 K, and an interstellar H$_{2}$:CO number ratio between 1 - 
2$\times$10$^{4}$, significantly smaller than the $<$6000 $M_{\earth}$ and 
$<$80 $M_{\earth}$ that we infer assuming $T_{ex}$ = 50 K and 100 K, 
respectively.

The non-detection of H$_{2}$ emission at mid-infrared wavelengths using 
\emph{Spitzer} IRS may be somewhat constraining for at least one object in our
sample. The expected H$_{2}$ emission line luminosity may be estimated from 
our inferred parent body masses (using the estimates in Table 5) assuming an 
interstellar gas:dust ratio and a gas excitation temperature, $T_{ex}$ = 
$T_{gr}$. If the system is a point source, then the total luminosity produced 
by $N(H_{2})$ H$_{2}$ molecules is
\begin{equation}
F = \frac{h \nu N(H_{2}) \chi_{u} A_{ul}}{4 \pi d^{2}}
\end{equation}
where $E = h \nu$  is the energy of the radiated photons, $\chi_{u}$ is the 
fraction of H$_{2}$ in level $u$, and $A_{ul}$ is the transition probability.
HD 113766 is a F3/F5V member of Lower Centaurus Crux in Scorpius-Centaurus,
with an estimated age of 16 Myr and a noteably large $L_{IR}/L_{*}$ = 0.015
\citep{chen05}. In our high-resolution mode sample, HD 113766 possesses the
largest parent body mass (0.1 $M_{\earth}$ or 260 times that mass in the
main asteroid belt) and the hottest single black body grain temperature, 
$T_{gr}$ = 330 K, making it the most likely object to possess $H_{2}$ S(0) and
S(1) emission. The approximation $T_{ex}$ = $T_{gr}$ is valid for optically 
thick, flared disks for $A_{V}$ $>$ 0.1 \citep{kd04} but invalid for debris 
disks around A-type stars where the disk gas and dust are too tenuous to heat 
the gas via gas-grain collisions \citep{kvz04}. For example, modeling of the 
gas around HR 4796A, an 8 Myr old A-type member of the TW Hydrae Association, 
located 67 pc away from the Sun, suggests that the gas, $T_{ex}$ = 65 K, is 
substantially cooler than the dust, $T_{gr}$ = 110 K \citep{ck04}. With these 
caveats, we estimate that the HD 113766 disk should possess H$_{2}$ S(0) and 
S(1) line fluxes of 2.8$\times$10$^{-15}$ ergs s$^{-1}$ cm$^{-2}$ and 
2.8$\times$10$^{-13}$ ergs s$^{-1}$ cm$^{-2}$, respectively, corresponding to 
an S(1) emission line flux $\sim$3$\times$ higher than our observed 3$\sigma$ 
upper limit.

\cite{gh04} have argued that [\ion{S}{1}] emission at 25.2 $\mu$m 
may be an excellent tracer for bulk gas in proto-planetary disks. Their 
model spectra of disks around intermediate-aged ($\sim$10$^{7}$ years) stars
predicts strong [\ion{S}{1}] emission, with line:continuum ratios $>$5\% in 
\emph{Spitzer} IRS high-resolution mode ($R$ $\sim$ 600), assuming that the 
disk sulfur abundance is similar to that observed in the diffuse interstellar 
medium (2.8$\times$10$^{-5}$) because co-spatial dust grains are too warm for 
volatiles to condense on them. However, the gas-phase sulfur abundance in 
intermediate-aged disks may be depleted relative to the diffuse interstellar
medium. The previously-identified 23.5 $\mu$m w\"{u}stite (FeO) emission 
feature detected in \emph{ISO} spectra of Herbig Ae stars may not be 
w\"{u}stite but pyrrhotite ([Fe,Ni]$_{1-x}$S). Keller et al. (2002) compare the
intrinsic strength of the 18 $\mu$m silicate feature to that of the 23.5 $\mu$m
pyrrhotite feature in the \emph{ISO} spectrum of HD 163496 and measure a 
silicon:sulfur ratio of 0.63, consistent with a solar abundance (0.52). They 
conclude that most if not all of the sulfur around HD 163296 resides in solid 
FeS grains. If the bulk of the circumstellar sulfur is located in grains, 
then [\ion{S}{1}] emission at 25.2 $\mu$m (or lack thereof) may not be a good 
bulk gas tracer.


\section{Conclusions}
We have obtained \emph{Spitzer Space Telescope} IRS spectra of 59 main 
sequence stars with spectral types B- through M-type and ages between 1 Myr 
and a few Gyr, that possess \emph{IRAS} 60 $\mu$m excess. We find that:

1. The majority of observed debris disks do not possess spectral features, 
suggesting that the grains are too cold and/or too large ($a$ $\geq$ 10 $\mu$m)
to produce spectral features. Detailed modeling of objects with spectral 
features requires the presence of large, warm amorphous silicates with 
$T_{gr}$ = 290 - 600 K, in addition to cool black body grains with $T_{gr}$ = 
80 - 200 K, and the presence of crystalline silicate mass ratios 0\% - 76\%.

2. The IRS spectra of debris disks (without spectral features) are generally
better fit using a single temperature black body than with a uniform disk. 
Stellar radiation pressure (in collisionally dominated systems), sublimation 
if the grains are icy, gas drag, and/or the presence of a perturbing body may 
contribute to the presence of inner holes in these disks. 

3. The peak in the distribution of estimated black body grain temperatures,
$T_{gr}$ = 110 - 120 K, suggesting that sublimation of icy grains may produce
the central clearings observed.

4. Since the parent body masses typically are less than the mass of the
Earth, it appears that planet formation efficiently consumes most of the mass
of the primordial disk.

\acknowledgements
We would like to thank K. Uchida for his assistance with data reduction and
our anonymous referee, S. Kenyon, E. Mamajek, J. Mould, I. Song, A. Speck, 
K. Stapelfeldt, and S. Strom for their helpful comments and suggestions. This 
material is based upon work supported by the National Aeronautics and Space 
Administration under Award No. NAS7-1407 and the California Institute of 
Technology.

\appendix
\section{Selected Source Notes}
{\bf HR 506} may possess a planet with a period of 1040 days, a minimum 
($M \sin i$) mass of 0.91 $M_{Jup}$, and a doppler velocity semi-amplitude of 
18 m s$^{-1}$; the inferred eccentricity and semi-major axis for the planetary
orbit are $e$ = 0.18 and $a$ = 2.1 AU (Mayor et al. 2003). If HR 506 possesses
black body grains at a distance $D$ = 2.1 AU from the star, then they are 
expected to have a grain temperature $T_{gr}$ = 220 K, assuming a stellar 
radius $R_{*}$ = 1.2 $R_{\sun}$ and an effective temperature $T_{eff}$ = 
6140 K, consistent with a stellar spectral type of F8V. However, the IRS excess
continuum emission observed toward this star is better fit by a black
body with a cooler grain temperature $T_{gr}$ = 70 K, suggesting that 
planetesimals at similar distances have already been removed. The black body 
distance for $T_{gr}$ = 70 K grains is $D$ = 21 AU. Therefore, the observed 
infrared emitting dust and parent body population are located at significantly 
larger distances than the radial velocity planet. A second planet in this 
system is probably needed to dynamically stir the parent bodies because the 
observed radial velocity planet is located too far away to generate strong 
orbital resonances (Bryden et al. 2000; Beichman et al. 2006).

{\bf $\tau$ Cet} is a 7.2 Gyr old G8V star (Lachaume et al. 1999) that 
possesses \emph{IRAS} and JCMT SCUBA excesses that are well fit by a $T_{gr}$ 
= 60 K modified black body (Greaves et al. 2004). However, the thermal emission
from this cold dust population is not well detected at IRS wavelengths 
($\lambda$ $<$ 35 $\mu$m). The slope of the 20 $\mu$m - 35 $\mu$m IRS spectrum 
is expected to be $\sim$4\% higher assuming the Greaves et al. (2004) model. 
The difficulty in detecting thermal emission from this population of 
circumstellar grains suggests that the IRS is sensitive to grains with dust 
temperature, $T_{gr}$ $>$ 60 K.

{\bf HR 3927} is a A0V star with an estimated age of 50 Myr. Its \emph{Spitzer}
IRS spectrum possesses a 10 $\mu$m feature that may be modeled using solid
amorphous olivine and crystalline forsterite grains with $a$ = 3.1 and 8
$\mu$m, respectively, larger than the blow-out size for both species, 
$T_{gr}$ = 290 K, and a crystalline silicate fraction of 38\%. The residual 
continuum emission may be modeled using black body grains with a lower grain 
temperature, $T_{gr}$ = 80 K. The presence of a cooler black body continuum in 
addition to the hot silicate grains suggests that this system possess two 
planetesimal belts in analogy with the asteroid and Kuiper belts in our solar 
system. The \emph{IRAS} (not color-corrected) 60 $\mu$m flux of 0.69 Jy is 
somewhat higher than the 0.34 Jy inferred from our model; however, the 
presence of a second source in the beam can not be ruled out without 
\emph{Spitzer} MIPS images. If HR 3927 does possess a larger 60 $\mu$m excess 
than predicted by our model, then it may also possess an additional (or third) 
population of cold dust with $T_{gr}$ $<$ 60 K.   

{\bf $\eta$ Crv} is a F2V star with an estimate age of 1 Gyr based on its
x-ray activity. This system may possess three populations of dusty debris. The 
coldest population with $T_{gr}$ = 40$\pm$5 K has been detected at 
submillimeter wavelengths where the system is resolved at all position angles 
at 450 $\mu$m with an elongation at a position angle of 
130$\arcdeg$$\pm$10$\arcdeg$ and at 850 $\mu$m with a radius of 100 AU
(Wyatt et al. 2005). A second component is detected in \emph{IRAS} photometry;
self-consistent modeling of the \emph{IRAS} and submillimeter photometry 
suggest that this population possesses a grain temperature $T_{gr}$ = 
370$\pm$60 K (Wyatt et al. 2005). Detailed fits to the \emph{Spitzer} IRS 
5.5 - 35 $\mu$m spectra suggest that this emission is produced by amorphous 
olivine, and crystalline forsterite and enstatite features with $T_{gr}$ = 
360 K and a crystalline silicate fraction of 31\%, and black body grains with a
temperature $T_{gr}$ = 120 K. The detection of three planetesimal belts may 
require the presence of at least two perturbing planets because three belts are
unlikely to be located at the resonances of a single planet. Detailed dynamical
models of this system are required to determine whether $\eta$ Crv possesses 
multiple planets.

{\bf HD 113766} is a binary member of Lower Centaurus Crux in the Sco-Cen OB 
Association with an estimated age of $\sim$16 Myr (Mamajek et al. 2002). The 
secondary lies 1.3$\arcsec$ (or 170 AU) away from the F3/F5 primary. HD 113766 
possesses an extremely high fractional infrared luminosity $L_{IR}/L_{*}$ = 
1.5$\times$10$^{-2}$ and MIPS 24 $\mu$m and 70 $\mu$m fluxes that are well-fit 
with a $T_{gr}$ = 330 K black body, suggestive of debris at terrestrial planet 
temperatures (Chen et al. 2005). Published ground-based ESO TIMMI2 spectra 
suggest that the 10 $\mu$m feature is dominated by crystalline silicate 
(forsterite) and large, amorphous silicates; SiO$_{2}$, which is correlated 
with the presence of forsterite in Herbig AeBe \emph{ISO} spectra, is not 
detected (Schutz et al. 2005). We use crystalline forsterite in addition to 
amorphous carbon, olivine, and pyroxene, and a single temperature black body to
fit not just the 10 $\mu$m feature but also the 5.5 $\mu$m - 35 $\mu$m IRS 
spectrum and infer a crystalline silicate fraction of 4.1\%. The presence of a 
cooler black body continuum ($T_{gr}$ = 200 K) in addition to the hot silicate 
and carbon grains ($T_{gr}$ = 600 K) suggests that this system possess two 
planetesimal belts in analogy with the asteroid and Kuiper belts in our solar 
system. The lack of spectral features due to sub-micron-sized grains can be 
explained if small particles are radiatively driven from the system by 
radiation pressure.

{\bf $\mu$ Ara} possess a planet with a period of 743 days, a minimum 
($M \sin i$) mass of 1.97 $M_{Jup}$, and a doppler velocity semi-amplitude of 
54 m s$^{-1}$; the inferred eccentricity and semi-major axis for the planetary
orbit are $e$ = 0.62 and $a$ = 1.65 AU (Butler et al. 2001). If the system
possessed black body grains at a distance $D$ = 1.65 AU from the star, then 
they would have a grain temperature $T_{gr}$ = 220 K, assuming a stellar 
radius $R_{*}$ = 0.99 $R_{\sun}$ and an effective temperature $T_{eff}$ = 
5830 K, consistent with a stellar spectral type of G3V. Grains with such
high temperatures are expected to produce a black body continuum that peaks 
near 25 $\mu$m in the center of the IRS bandpass, similar to that observed 
toward $\zeta$ Lep. However, no excess emission has been detected toward 
$\mu$ Ara at IRS wavelengths. One possible explanation for the lack of infrared
excess is that this system has already destroyed any planetesimal belt it may 
have possessed by its estimated age of 6.2 Gyr (Lachaume et al. 1999).

%
{\bf HR 7012} is a A5IV-V member of the $\beta$ Pic moving group with an 
estimated age of 12 Myr (Zuckerman et al. 2001). Its \emph{Spitzer} IRS
spectrum possesses a strong 10 $\mu$m silicate feature that may be modeled
using large amorphous olivine and pyroxene grains with $a$ = 5.0 and 1.1
$\mu$m, respectively, and $T_{gr}$ = 520 K and sub-micron-sized crystalline
enstatite and cristobalite. Cristobalite is used to fit emission in the
shoulder of the silicate feature at $\lambda$ $<$ 8.7 $\mu$m, however
other materials such as opal (hydrated silica) may also provide acceptable
fits. We estimate that this system possess a crystalline silicate fraction of 
76\% however our model produces extra emission at 10.5 and 19.5 $\mu$m that 
may indicate that the fraction of crystalline pyroxene used is too high. The 
presence of small sub-$\mu$m-sized grains inferred from our fit to the HR 7012 
spectrum suggest that this system may have experienced a recent collision.

{\bf $\eta$ Tel} is a binary member of the $\beta$ Pic moving group with 
an estimated age of 12 Myr (Zuckerman et al. 2001). The M7/8V secondary
located at a distance of 4$\arcsec$ (or 200 AU) away from the A0V primary was
detected via common proper motion studies (Lowrance et al. 2000).  The
$\eta$ Tel \emph{Spitzer} IRS spectrum possesses a 10 $\mu$m feature that may 
be modeled using solid amorphous olivine grains with $a$ = 3.0 $\mu$m, larger 
than the blow-out size, and $T_{gr}$ = 370 K. The residual continuum emission 
may be modeled using black body grains with a lower temperature, $T_{gr}$ 
= 115 K. The presence of a cooler black body continuum in addition to the hot 
silicate grains suggests that this system possess two planetesimal belts in 
analogy with the asteroid and Kuiper belts in our solar system. This spectrum 
does not appear to possess crystalline forsterite or enstatite features despite
the young age of the system. For comparison, HR 7012 (another A-type member of 
the $\beta$ Pic moving group) has a crystalline silicate fraction of 76\%.


\clearpage
\pagestyle{empty}
\begin{deluxetable}{cccccccccccrrrc}
\rotate
\tablecaption{Stellar Properties}
\tablehead{
    \omit &
    \omit &
    \omit &
    \omit &
    \omit &
    \omit &
    \omit &
    \omit &
    \omit &
    \colhead{This Work} &
    \colhead{Literature} &
    \colhead{25 $\mu$m} &
    \colhead{60 $\mu$m} &
    \colhead{100 $\mu$m} &
    \colhead{Excess}\\
    \colhead{HR} &
    \colhead{HD} &
    \colhead{Name} &
    \colhead{Spectral} &
    \colhead{Distance} &
    \colhead{$v \sin i$} &
    \colhead{$A_{V}$} &
    \colhead{$T_{eff}$} &
    \colhead{$\log g$} &
    \colhead{Age} &
    \colhead{Age} &
    \colhead{Excess} & 
    \colhead{Excess} & 
    \colhead{Excess} & 
    \colhead{Reference}\\
    \omit &
    \omit &
    \omit &
    \colhead{Type} &
    \colhead{(pc)} &
    \colhead{(km/sec)} &
    \colhead{(mag)} &
    \colhead{(K)} &
    \omit &
    \colhead{(Gyr)} &
    \colhead{(Gyr)} &
    \colhead{(Jy)}  &
    \colhead{(Jy)}  &
    \colhead{(Jy)}  &
    \omit\\
}
\tablewidth{0pt}
\tablecolumns{15}
\startdata
  123 &   2772 & $\lambda$ Cas$^{\dagger}$ & B8Vn   & 109 & 220 & 0.03 & 13290 & 3.946 & 0.11  & 0.01$^{a}$          & 0.00 & 0.96 & 2.44 & 1 \\ 
  333 &   6798 &                           & A3V    &  83 & 190 & 0.00 & 10360 & 4.243 & 0.11  &      ...            & 0.00 & 0.34 & 0.00 & 1 \\
  451 &   9672 & 49 Cet                    & A1V    &  61 & 195 & 0.00 &  9970 & 4.369 & 0.05  & 0.02$^{i}$          & 0.33 & 1.99 & 1.90 & 1,3,4,6 \\
  493 &  10476 & 107 Psc                   & K1V    &  7  &  10 & 0.00 &  5390 & 4.878 & ...   &      ...            & 0.10 & 0.11 & 0.38 & 1 \\ 
  506 &  10647 &                           & F9V    &  17 &   5 & 0.00 &  6260 & 4.598 & 0.3   & 0.3, 4.8$^{d,i}$    & 0.14 & 0.81 & 0.00 & 1,3 \\ 
  509 &  10700 & $\tau$ Cet                & G8V    &   4 &   7 & 0.00 &  6000 & 5.288 & ...   & 7.2$^{c}$           & 0.06 & 0.08 & 0.42 & 1,3 \\
  664 &  14055 & $\gamma$ Tri              & A1Vnn  &  36 & 230 & 0.00 & 10540 & 4.172 & 0.17  & 0.16$^{i}$          & 0.21 & 0.81 & 0.83 & 1 \\ 
  ... &  16157 & CC Eri                    & M0     &  12 & ... & 0.00 &  3900 & ...   & ...   &      ...            & 0.16 & 0.10 & 0.21 & 3 \\
  818 &  17206 & $\tau^1$ Eri              & F5/F6V &  14 &  25 & 0.00 &  6480 & 4.500 & 0.3   & 3.5$^{d}$           & 0.17 & 0.89 & 3.65 & 1 \\ 
  919 &  18978 & $\tau^3$ Eri              & A4V    &  26 & 120 & 0.00 &  8610 & 4.110 & 0.5   &      ...            & 0.02 & 0.04 & 0.15 & 1 \\
  963 &  20010 & $\alpha$ For              & F8IV   &  14 &  15 & 0.00 &  6360 & 4.169 & 2.7   & 4.3$^{d}$           &-0.02 & 0.11 & 0.00 & 1,6 \\ 
 1082 &  21997 &                           & A3IV/V &  74 &  60 & 0.00 &  9000 & 4.349 & 0.15  & 0.1$^{i}$           & 0.00 & 0.58 & 0.00 & 1 \\
 1338 &  27290 & $\gamma$ Dor              & F4III  &  20 &  65 & 0.00 &  7290 & 4.260 & 0.40  &      ...            & 0.05 & 0.21 & 0.21 & 1,3,6 \\ 
 1570$^{\dagger}$ &  31295 &               & A0V    &  37 & 110 & 0.43 &  9450 & 4.212 & 0.23  & 0.01, 0.1$^{e,i}$   & 0.00 & 0.42 & 0.00 & 1,4 \\ 
 1686$^{\dagger}$ &  33564 &               & F6V    &  21 &  10 & 0.00 &  6430 & 4.238 & 2.0   & 3.0$^{d}$           &-0.04 & 0.23 & 0.00 & 1 \\ 
 1705 &  33949 & $\kappa$ Lep$^{\dagger}$  & B7V    & 172 & 125 & 0.00 & 12740 & 3.497 & 0.12  &        ...          & 0.13 & 0.41 & ...  & 3,4 \\
 1839 &  36267 & 32 Ori$^{\dagger}$        & B5V    &  89 & 190 & 0.00 & 16430 & 4.441 & 0.001 &        ...          & 0.00 & 0.55 & 0.00 & 1 \\ 
 1998 &  38678 & $\zeta$ Lep               & A2Vann &  22 & 245 & 0.00 &  9910 & 4.213 & 0.18  & 0.23$^{g}$          & 0.68 & 0.40 & $<$0.11 & 1,2,3 \\
 2015 &  39014 & $\delta$ Dor              & A7V    &  44 & 170 & 0.00 &  8360 & 3.734 & 0.59  & 0.49, 0.54$^{e,g}$  & 0.04 & 0.45 & 1.11 & 1,2 \\ 
 2124$^{\dagger}$ &  40932 &               & A2V    &  47 &  18 & 0.47 &  8350 & 3.969 & 0.67  & 0.69$^{g}$          & 0.63 & 3.20 & 2.75 & 1 \\ 
 2161 &  41814 & XZ Lep                    & B3V    & 496 &  35 & 0.12 & 17890 & 4.197 & 0.017 &        ...          & 0.00 & 0.57 & 0.00 & 1 \\
 2483 &  48682 & $\psi^5$ Aur$^{\dagger}$  & G0V    &  17 &   5 & 0.00 &  6350 & 4.608 &  ...  & 4.5$^{d}$           & 0.11 & 0.43 & 0.53 & 1 \\ 
  ... &  53143 &                           & K1V    &  18 &   4 & 0.00 &  5000 & ...   & ...   & 0.3, 0.97$^{f,i}$   & 0.02 & 0.14 & 0.66 & 3 \\
 3220 &  68456 &                           & F5V    &  21 &  15 & 0.00 &  6600 & 4.259 & 1.7   & 2.6$^{d}$           & 0.04 & 1.76 & 3.05 & 1 \\ 
 3314 &  71155 &                           & A0V    &  38 & 120 & 0.00 & 10190 & 4.190 & 0.19  & 0.17, 0.24$^{e,g}$  &-0.03 & 0.57 & ...  & 2 \\
 3485 &  74956 & $\delta$ Vel              & A1V    &  24 & 150 & 0.00 &  9820 & 3.904 & 0.35  & 0.39$^{g}$          & 0.03 & 0.29 & $<$0.02 & 1,2,6 \\
  ... &   ...  & FI Vir                    & M4     &   3 & ... & 0.00 &  3400 & ...   & ...   &        ...          & ...  & 0.12 & 0.22 & 1 \\ 
 3862 &  84117 &                           & G0V    &  15 &   6 & 0.00 &  6290 & 4.338 & 1.9   & 4.6$^{d}$           & 0.12 & 0.13 & 0.24 & 1 \\ 
 3927 &  86087 &                           & A0V    &  98 & ... & 0.00 & 10090 & 4.400 & 0.05  &        ...          & 0.00 & 0.68 & 0.00 & 1 \\
  ... &  95086 &                           & A8III  &  92 & ... & 0.00 &  7500 & ...   & ...   & 0.016$^{a}$         & $<$0.25 & 0.60 & $<$0.80 & 3 \\ 
 4295 &  95418 & $\beta$ UMa               & A1V    &  24 &  32 & 0.00 &  9790 & 3.881 & 0.34  & 0.36$^{g}$          & 0.24 & 0.43 & 0.00 & 1,2 \\
 4534 & 102647 & $\beta$ Leo               & A3V    &  11 & 110 & 0.00 &  9020 & 4.293 & 0.05  & 0.05$^{g}$          & 0.41 & 0.77 & 0.63 & 1,2,6 \\
 4732 & 108257 & G Cen$^{\dagger}$         & B3Vn   & 123 & 245 & 0.16 & 17930 & 4.140 & 0.025 & 0.016$^{a}$         & 0.44 & 0.88 & 0.00 & 1 \\ 
 4775 & 109085 & $\eta$ Crv                & F2V    &  18 &  92 & 0.00 &  6890 & 4.28  & ...   & 1.0$^{j}$           & 2.24 & 0.77 & 0.31 & 3,5 \\
  ... & 110058 &                           & A0V    & 100 & ... & 0.40 &  9500 & ...   & ...   & 0.016$^{a}$         & 0.30 & 0.51 & $<$1.00 & 3 \\ 
  ... & 113766$^{\dagger}$ &               & F3/F5V & 131 & ... & 0.05 &  6870 & 4.360 & ...   & 0.016$^{a}$         & 1.80 & 0.65 & $<$1.00 & 3 \\ 
 5236 & 121384 &                           & G6IV-V &  38 & ... & 0.00 &  5670 & 5.070 & ...   & 4.1$^{d}$           & 0.00 & 0.34 & $<$10.2 & 6 \\ 
 5351 & 125162 & $\lambda$ Boo             & A0p    &  30 & 110 & 0.00 &  9310 & 4.193 & 0.27  & 0.18, 0.31$^{e,g}$  & 0.09 & 0.45 & ...  & 4 \\
 5447 & 128167 & $\sigma$ Boo              & F2V    &  15 &  10 & 0.00 &  6830 & 4.408 & 1.4   & 1.0, 1.7$^{c,d,i}$  & 0.06 & 0.09 & $<$0.06 & 1 \\ 
 5671 & 135382 & $\gamma$ Tra              & A1V    &  56 & 200 & 0.03 & 10060 & 3.244 & 0.17  & 0.26$^{e}$          & 0.06 & 0.07 & 1.51 & 2 \\ 
 5793 & 139006 & $\alpha$ CrB              & A0V    &  23 & 132 & 0.00 & 10180 & 3.949 & 0.30  & 0.31$^{g}$          & 0.31 & 0.50 & 0.00 & 1,2 \\
  ... & 139664 &                           & F5IV-V &  18 &  87 & 0.02 &  6900 & 4.450 & 0.48  &0.2,1.1,1.6$^{c,d,i}$& 0.33 & 0.54 & $<$2.35 & 6 \\ 
 5933 & 142860 & 41 Ser                    & F6IV   &  11 &  10 & 0.00 &  6430 & 4.348 & 1.6   & 3.2, 3.3$^{c,d}$    &-0.08 & 0.24 & 0.00 & 1 \\ 
  ... & 146897 &                           & F2/F3V & 132 & ... & 0.31 &  6750 & ...   & ...   & 0.005$^{a}$         & $<$0.6 & 0.73 & $<$3.05 \\ 
 6168 & 149630 & $\sigma$ Her              & B9V    &  93 & 285 & 0.00 & 11440 & 3.685 & 0.17  & 0.23$^{b}$          & 0.02 & 0.22 & 0.00 & 1,2,4 \\ 
 6297 & 153053 &                           & A5IV-V &  51 & ... & 0.22 &  8070 & 4.030 & 0.42  &        ...          & 1.03 & 1.81 & ...  & 2 \\ 
 6486 & 157792 &                           & A3m... &  26 &  68 & 0.00 &  7660 & 4.222 & 0.24  & 0.74$^{c}$          & 0.15 & 1.02 &11.48 & 1 \\ 
 6532$^{\dagger}$ & 159082 &               & B9     & 152 &  20 & 0.25 & 11210 & 4.019 & 0.2   &        ...          & 0.00 & 0.69 & 1.28 & 1 \\
 6533 & 159139 & 78 Her                    & A1V    &  84 & 260 & 0.00 & 10980 & 4.305 & 0.05  &        ...          & 0.00 & 0.46 & 1.23 & 1 \\ 
 6585 & 160691 & $\mu$ Ara                 & G3IV-V &  15 & ... & 0.00 &  5500 & ...   & ...   & 6.2$^{c}$           & 0.22 & 0.08 & $<$0.96 & 1 \\ 
 6629 & 161868 & $\gamma$ Oph              & A0V    &  29 & 212 & 0.00 & 10200 & 4.189 & 0.19  & 0.18$^{g}$          & 0.17 & 1.23 & 0.00 & 1,2,4 \\
 6670 & 162917 &                           & F4IV-V &  31 &  20 & 0.00 &  6670 & 4.369 & 1.6   & 1.4$^{d}$           & $<$0.09 & 0.52 & ...  & 4 \\ 
 7012 & 172555 &                           & A5IV-V &  29 & 175 & 0.00 &  8550 & 4.377 & 0.05  & 0.012$^{h}$         & 0.61 & 1.92 & ...  & 2 \\ 
 7329 & 181296 & $\eta$ Tel$^{\dagger}$    & A0Vn   &  48 & 420 & 0.12 & 10180 & 4.893 & 0.05  & 0.012$^{h}$         & 0.35 & 0.51 & 0.00 & 1,3 \\ 
  ... & 181327 &                           & F5/F6V &  51 & ... & 0.00 &  6560 & 4.510 & 1.4   & 0.012, 1.3$^{d,h}$  & 0.20 & 2.00 & 1.90 & 3 \\
  ... & 191089 &                           & F5V    &  54 & ... & 0.00 &  6540 & 4.420 & 1.6   & $<$0.1, 3.0$^{d,i}$ & 0.34 & 0.71 & 0.33 & 3 \\ 
  ... & 200800 &                           & A3Vm...& 127 & ... & 0.09 &  8650 & ...   & ...   &        ...          & 0.31 & 0.15 & 0.34 & 3 \\
 8799 & 218396 &                           & A5V    &  40 &  40 & 0.34 &  7410 & 4.163 & 0.59  & 0.03, 0.73$^{g,i}$  & $<$0.19 & 0.45 & ...  & 4 \\
  ... & 221354 &                           & K2V    &  17 & ... & 0.00 &  5350 & 4.270 & ...   & 1.5$^{f}$           & 0.09 & 0.88 & $<$20.5 & 6 \\ 
\enddata
\tablenotetext{\dagger}{binary system}
\tablerefs{(a) de Zeeuw et al. 1999; (b) Grosbol 1978; (c) Lachaume et al. 1999; 
  (d) Nordstrom et al. 2004; (e) Paunzen et al. 1997; (f) Song et al. 2000; 
  (g) Song et al 2001; (h) Zuckerman et al. 2001; (i) Zuckerman \& Song 2004
  (j) Wyatt et al. 2005}
\tablerefs{(1) Backman \& Paresce 1993; (2) Cote 1987;
  (3) Mannings \& Barlow 1998; (4) Sadakane \& Nishida 1986;
  (5) Sylvester et al. 1996; (6) Walker \& Wolstencroft 1988}
\end{deluxetable}
\clearpage
\pagestyle{plaintop}


\begin{deluxetable}{l|cccc|cccc|c}
\rotate
\tablecaption{IRS Photometry in 8.5-13 and 30-34 $\mu$m Bands}
\tablehead{
    \omit &
    \multicolumn{4}{c}{8.5 - 13 $\mu$m} &
    \multicolumn{4}{c}{30 - 34 $\mu$m} \\
    \omit &
    \colhead{Measured} &
    \colhead{Predicted} &
    \omit &
    \omit &
    \colhead{Measured} &
    \colhead{Predicted} &
    \omit &
    \omit &
    \omit \\
    \colhead{Name} &
    \colhead{Flux} &
    \colhead{Flux} &
    \colhead{Excess} &
    \colhead{Fractional} &
    \colhead{Flux} &
    \colhead{Flux} &
    \colhead{Excess} &
    \colhead{Fractional} &
    \colhead{\emph{Spitzer}} \\
    \omit &
    \colhead{(mJy)} &
    \colhead{(mJy)} &
    \colhead{(mJy)} &
    \colhead{Excess} &
    \colhead{(mJy)} &
    \colhead{(mJy)} &
    \colhead{(mJy)} &
    \colhead{Excess} &
    \colhead{AOR Key} \\
}
\tablewidth{0pt}
\tablecolumns{10}
\startdata
$\lambda$ Cas             &  341$\pm$0.8   &  350   &   -9.3 & -0.027  &   55.9$\pm$1.4 &  37.5 &  19.2 &  0.524 & 3575040 \\
HR 333                    &  227$\pm$0.3   &  212   &   14.7 &  0.069  &  161$\pm$2     &  23.4 & 138   &  6.14  & 3553536 \\
49 Cet                    &  270$\pm$0.6   &  218   &   52.7 &  0.242  &  426$\pm$1     &  23.9 & 402   & 16.8   & 4928768 \\ 
107 Psc$^{\ddagger}$      & 1860$\pm$0.5   & 1880   &  -29.1 & -0.015  &  201$\pm$1     & 196   &  -5.3 & -0.026 & 3554048 \\ 
HR 506                    &  747$\pm$0.9   &  773   &  -26.4 & -0.034  &  197$\pm$2     &  84.4 & 114   &  1.36  & 3553792 \\
$\tau$ Cet$^{\ddagger}$   & 7470$\pm$10    & 7710   & -237   & -0.031  &  786$\pm$3     & 829   & -42.8 & -0.052 & 4932352 \\
$\gamma$ Tri              &  951$\pm$1     &  923   &   27.4 &  0.030  &  327$\pm$2     &  97.1 & 229   &  2.34  & 3554304 \\ 
CC Eri$^{\ddagger}$       &  439$\pm$0.6   &  558   & -119   & -0.213  &   55.6$\pm$1.1 &  59.9 &  -5.7 & -0.093 & 3554560 \\
$\tau^1$ Eri$^{\ddagger}$ & 1740$\pm$2     & 1780   &  -40.3 & -0.023  &  186$\pm$2     & 199   &  -4.9 & -0.026 & 3554816 \\ 
$\tau^3$ Eri              & 1270$\pm$1     & 1250   &   29.3 &  0.024  &  196$\pm$1     & 132   &  63.3 &  0.477 & 4931840 \\ 
$\alpha$ For$^{\ddagger}$ & 3410$\pm$4     & 3450   &  -38.7 & -0.011  &  413$\pm$5     & 378   &  28.3 &  0.074 & 3555072 \\ 
HR 1082                   &  125$\pm$0.3   &  123   &    1.7 &  0.014  &  117$\pm$2     &  12.9 & 104   &  7.96  & 3555328 \\ 
$\gamma$ Dor$^{\ddagger}$ & 1540$\pm$2     & 1580   &  -36.2 & -0.023  &  194$\pm$2     & 170   &  24.8 &  0.147 & 3555584 \\ 
HR 1570                   &  635$\pm$0.9   &  663   &  -29.4 & -0.044  &  175$\pm$2     &  69.0 & 105   &  1.50  & 3555840 \\ 
HR 1686$^{\ddagger}$      &  889$\pm$1     &  936   &  -46.9 & -0.050  &  106 $\pm$2    & 101   &   6.1 &  0.061 & 3556096 \\ 
$\kappa$ Lep$^{\ddagger}$ &  521$\pm$0.6   &  517   &    4.1 &  0.008  &   73.1$\pm$1.3 &  53.1 &  19.2 &  0.356 & 3577600 \\
32 Ori$^{\ddagger}$       &  468$\pm$0.5   &  492   &  -23.6 & -0.048  &   50.4$\pm$1.7 &  47.1 &  -0.6 & -0.012 & 3556352 \\
$\zeta$ Lep               & 1950$\pm$0.6   & 1710   &  232   &  0.136  &  886$\pm$4     & 182   & 704   &  3.87  & 4932864 \\
$\delta$ Dor$^{\ddagger}$ & 1210$\pm$1     & 1240   &  -33.1 & -0.027  &  135$\pm$2     & 131   &   3.2 &  0.024 & 3556608 \\
HR 2124$^{\ddagger}$      & 1260$\pm$2     & 1350   &  -95.1 & -0.070  &  203$\pm$5     & 144   &  54.2 &  0.365 & 3556864 \\
XZ Lep$^{\ddagger}$       &   48.1$\pm$0.3 &   45.3 &    2.7 &  0.061  &    5.5$\pm$2.1 &   4.6 &   0.8 &  0.168 & 3585536 \\
$\psi^5$ Aur$^{\ddagger}$ & 1010$\pm$1     & 1045   &  -35.8 & -0.034  &  140$\pm$2     & 111   &  28.4 &  0.255 & 3557120 \\
HD 53143                  &  380$\pm$0.4   &  395   &  -14.6 & -0.037  &   82.0$\pm$1.1 &  41.6 &  39.3 &  0.919 & 3557632 \\
HR 3220$^{\ddagger}$      & 1260$\pm$2     & 1260   &   -7.1 & -0.006  &  143$\pm$2     & 137   &   7.7 &  0.057 & 3557888 \\
HR 3314                   &  987$\pm$0.8   &  978   &    9.5 &  0.010  &  281$\pm$2     & 103   & 177   &  1.71  & 3558144 \\
$\delta$ Vel$^{\ddagger}$ & 6850$\pm$7     & 6810   &   35.1 &  0.005  &  766$\pm$4     & 721   &  44.6 &  0.062 & 4930816 \\
FI Vir$^{\ddagger}$       &  343$\pm$0.4   &  348   &   -5.6 & -0.016  &   37.5$\pm$1.4 &  41.5 &  -2.2 & -0.054 & 3559168 \\ 
HR 3862$^{\ddagger}$      & 1310$\pm$1     & 1370   &  -66.0 & -0.048  &  140$\pm$2     & 146   &  -7.1 & -0.048 & 3558400 \\
HR 3927                   &  195$\pm$0.4   &  177   &   18.1 &  0.102  &  187$\pm$2     &  18.8 & 168   &  8.94  & 3558656 \\
HD 95086                  &   75.3$\pm$0.3 &   69.0 &   12.5 &  0.199  &  114$\pm$2     &   7.6 & 107   & 14.6   & 3558912 \\
$\beta$ UMa               & 4110$\pm$5     & 4160   &  -48.8 & -0.012  &  801$\pm$3     & 440   & 361   &  0.820 & 4930304 \\
$\beta$ Leo               & 6390$\pm$7     & 6370   &   18.3 &  0.003  & 1410$\pm$6     & 676   & 736   &  1.09  & 4929793 \\
G Cen$^{\ddagger}$        &  259$\pm$0.5   &  268   &   -7.9 & -0.029  &   34.7$\pm$2.2 &  27.3 &   7.1 &  0.269 & 3579392 \\
$\eta$ Crv                & 1890$\pm$3     & 1610   &  276   &  0.171  &  476$\pm$1     & 179   & 197   &  1.10  & 3559424 \\
HD 110058                 &   55.3$\pm$0.3 &   40.3 &   15.0 &  0.372  &  380$\pm$2     &   4.3 & 376   & 88.1   & 3579648 \\
HD 113766                 & 1870$\pm$2     &   86.2 & 1780   & 20.650  &  988$\pm$4     &   9.6 & 978   &102     & 3579904 \\
HR 5236$^{\ddagger}$      &  931$\pm$0.5   &  901   &   28.9 &  0.032  &  105$\pm$0.9   &  96.1 &   7.6 &  0.078 & 3559680 \\
$\lambda$ Boo             &  936$\pm$0.9   &  937   &   -0.5 & -0.001  &  272$\pm$2     & 100   & 173   &  1.75  & 3559936 \\
$\sigma$ Boo$^{\ddagger}$ & 1510$\pm$1     & 1540   &  -29.8 & -0.019  &  172$\pm$2     & 164   &   7.4 &  0.045 & 3560192 \\
$\gamma$ Tra$^{\ddagger}$ & 3090$\pm$2     & 3190   &  -99.8 & -0.031  &  359$\pm$5     & 345   &   9.2 &  0.026 & 3560448 \\ 
$\alpha$ CrB              & 4860$\pm$2     & 4830   &   30.0 &  0.006  &  988$\pm$4     & 511   & 476   &  0.932 & 4929280 \\
HD 139664                 & 1260$\pm$1     & 1250   &    7.7 &  0.006  &  223$\pm$2     & 135   &  89.2 &  0.666 & 3560704 \\
41 Ser$^{\ddagger}$       & 3160$\pm$3     & 3320   & -159   & -0.048  &  364$\pm$4     & 368   &  -5.4 & -0.015 & 3561216 \\
HD 146897                 &   34.7$\pm$0.3 &   27.0 &    7.6 &  0.283  &  329$\pm$3     &   2.9 & 327   &113     & 3581696 \\
$\sigma$ Her$^{\ddagger}$ &  750$\pm$0.8   &  797   &  -46.9 & -0.059  &   87.0$\pm$3.5 &  85.2 &   2.9 &  0.034 & 3561472 \\
HR 6297                   &  271$\pm$0.3   &  287   &  -16.8 & -0.058  &   80.4$\pm$2.2 &  31.1 &  49.9 &  1.64  & 3561728 \\
HR 6486                   & 1470$\pm$2     & 1520   &  -59.9 & -0.039  &  310$\pm$5     & 168   & 142   &  0.846 & 3562240 \\
HR 6532                   &  106$\pm$0.3   &  102   &    4.0 &  0.039  &   40.2$\pm$2.0 &  10.5 &  29.4 &  2.74  & 3582720 \\
78 Her                    &  208$\pm$0.4   &  208   &    0.9 &  0.004  &   47.9$\pm$12.7&  21.9 &  25.9 &  1.18  & 3562496 \\
$\mu$ Ara$^{\ddagger}$    & 1400$\pm$1     & 1490   &  -91.2 & -0.061  &  146$\pm$3     & 172   & -14.9 & -0.093 & 3562753 \\
$\gamma$ Oph              & 1340$\pm$2     & 1320   &   19.1 &  0.014  &  729$\pm$3     & 134   & 589   &  4.23  & 4931328 \\
HR 6670                   &  442$\pm$0.5   &  456   &  -13.5 & -0.030  &   78.2$\pm$1.9 &  47.7 &  29.4 &  0.602 & 3563008 \\
HR 7012                   & 1410$\pm$1     &  658   &  753   &  1.144  &  723$\pm$5     &  73.3 & 649   &  8.85  & 3563264 \\
$\eta$ Tel                &  468$\pm$0.7   &  333   &  135   &  0.405  &  469$\pm$2     &  35.1 & 433   & 12.3   & 3563776 \\
HD 181327                 &  161$\pm$0.3   &  154   &    7.1 &  0.046  &  555$\pm$2     &  15.8 & 539   & 32.6   & 3564032 \\
HD 191089                 &  134$\pm$0.4   &  131   &    3.0 &  0.023  &  363$\pm$4     &  14.2 & 349   & 24.9   & 3564288 \\
HD 200800                 &   47.2$\pm$0.3 &   43.0 &    4.2 &  0.097  &    9.5$\pm$5.9 &   4.6 &   5.0 &  1.09  & 3582976 \\
HR 8799                   &  290$\pm$0.4   & 280    &   10.5 &  0.038  &   72.1$\pm$2.1 &  30.1 &  42.3 &  1.42  & 3565568 \\
HD 221354$^{\ddagger}$    &  441$\pm$0.5   & 469    &  -28.2 & -0.060  &   43.4$\pm$1.1 &  49.4 &  -7.1 & -0.140 & 3565824 \\
\enddata
\tablenotetext{\ddagger}{stars without strong IRS excess}
\end{deluxetable}
\clearpage
\pagestyle{plaintop}


\begin{deluxetable}{lccccc}
\tablecaption{Spectral Feature Fitting Parameters}
\small
\tablehead{
\omit &
\colhead{HR 3927} &
\colhead{$\eta$ Crv} &
\colhead{HD 113766} &
\colhead{HR 7012} &
\colhead{$\eta$ Tel} \\
}
\tablewidth{0pt}
\tablecolumns{6}
\startdata
\cutinhead{Black Body Components}
$T_{bb1}$ (K)                  &   80 &  120 &  200 &  200 &  115 \\
$\Omega_{bb1}$ ($10^{-16}$ sr) &  350 &   80 &   68 &   42 &  160 \\
$T_{bb2}$ (K)                  &  ... &  ... &  ... &  ... &  370 \\
$\Omega_{bb1}$ ($10^{-16}$ sr) &  ... &  ... &  ... &  ... &  1.1 \\
\cutinhead{Temperatures}
$T_{silicate}$ (K)        &  290 &  360 &  600 &  520 &  370 \\
$T_{carbon}$ (K)          &  ... &  ... &  600 &  520 &  ... \\
Silicate $a_{min,o}$ ($\mu$m) & 3.1 & 0.54 & 1.4 & 1.1 & 2.4 \\
Carbon $a_{min,o}$ ($\mu$m)   & ... &  ... & 1.9 & 1.4 & ... \\
Silica $a_{min,o}$ ($\mu$m)   & ... &  ... & ... & 1.6 & ... \\
\cutinhead{Amorphous Olivine Properties}
Composition              & MgFeSi0$_{4}$ & MgFeSi0$_{4}$ & MgFeSi0$_{4}$ & MgFeSi0$_{4}$ & MgFeSi0$_{4}$ \\
Shape                    &       spheres &       spheres &       spheres &       spheres &       spheres \\
$a$ ($\mu$m)             &           3.1 &          3.5  &           1.5 &           5.0 &           3.0 \\
$f$ ($V_{vac}/V_{tot}$)  &           0   &          0.35 &           0   &           0.6 &           0   \\
$m$ ($10^{20}$ g)        &          24   &          7.8  &          82   &           4.4 &           8.0 \\
\cutinhead{Amorphous Pyroxene Properties}
Composition              & ... & ... & Mg$_{0.5}$Fe$_{0.5}$Si0$_{3}$ & Mg$_{0.8}$Fe$_{0.2}$Si0$_{3}$ & ... \\
Shape                    & ... & ... &                       spheres &                       spheres & ... \\
$a$ ($\mu$m)             & ... & ... &                           1.5 &                           1.1 & ... \\
$f$ ($V_{vac}/V_{tot}$)  & ... & ... &                            0  &                           0   & ... \\
$m$ ($10^{20}$ g)        & ... & ... &                           97  &                           2.4 & ... \\
\cutinhead{Forsterite Properties}
Composition              & Mg$_{1.9}$Fe$_{0.1}$Si0$_{4}$ & Mg$_{1.9}$Fe$_{0.1}$Si0$_{4}$ & Mg$_{2}$Si0$_{4}$ & ... & ... \\
Shape                    &                      spheres  &                       spheres &        nonspheres & ... & ... \\
$a$ ($\mu$m)             &                           8   &                          8    &        sub-$\mu$m & ... & ... \\
$f$ ($V_{vac}/V_{tot}$)  &                           0   &                          0    &              0    & ... & ... \\
$m$ ($10^{20}$ g)        &                          15   &                          0.47 &              7.7  & ... & ... \\
\cutinhead{Enstatite Properties}
Composition              & ... & MgSi0$_{3}$ & ... & Mg$_{0.7}$Fe$_{0.3}$Si0$_{4}$ & ... \\
Shape                    & ... &     spheres & ... &                    nonspheres & ... \\
$a$ ($\mu$m)             & ... &        1    & ... &                    sub-$\mu$m & ... \\
$f$ ($V_{vac}/V_{tot}$)  & ... &        0.4  & ... &                             0 & ... \\
$m$ ($10^{20}$ g)        & ... &        3    & ... &                            20 & ... \\
\cutinhead{Silica Properties}
Composition              & ... & ... & ... &   Cristobalite & ... \\
Shape                    & ... & ... & ... &           CDE2 & ... \\
$a$ ($\mu$m)             & ... & ... & ... & Rayleigh-limit & ... \\
$f$ ($V_{vac}/V_{tot}$)  & ... & ... & ... &              0 & ... \\
$m$ ($10^{20}$ g)        & ... & ... & ... &            1.5 & ... \\
\cutinhead{Amorphous Carbon Properties}
$a$ ($\mu$m)             & ... & ... &  1.9 & 1.5 & ... \\
$f$ ($V_{vac}/V_{tot}$)  & ... & ... &  0   & 0   & ... \\
$m$ ($10^{20}$ g)        & ... & ... & 55   & 0.7 & ... \\
\enddata
\end{deluxetable}
\clearpage
\pagestyle{empty}
\begin{deluxetable}{lccccccccccc}
\rotate
\small
\tablecaption{Dust Properties Inferred from Excess Continua}
\tablehead{
\omit &
\omit &
\omit &
\omit &
\omit &
\omit &
\omit &
\omit &
\omit &
\colhead{Black Body} &
\omit &
\colhead{Uniform Disk} \\
\colhead{Name} &
\colhead{$L_{*}$} &
\colhead{$M_{*}$} &
\colhead{$T_{gr}$} &
\colhead{$L_{IR}/L_{*}$} &
\colhead{$D$} &
\colhead{$<a>$} &
\colhead{$M_{dust}$} &
\colhead{$M_{10 cm}$} &
\colhead{Minimum} &
\colhead{$K2$} &
\colhead{Minimum} \\
\omit &
\colhead{($L_{\sun}$)} &
\colhead{($M_{\sun}$)} &
\colhead{(K)} &
\omit &
\colhead{(AU)} &
\colhead{($\mu$m)} &
\colhead{($M_{\earth}$)} &
\colhead{($M_{\earth}$)} &
\colhead{$\chi^{2}$} &
\colhead{(Jy/$\mu$m)} &
\colhead{$\chi^{2}$} \\
}
\tablewidth{0pt}
\tablecolumns{12}
\startdata
 $\lambda$ Cas  &   250   & 4.0 & 120$\pm$100 & 4.4$\times$10$^{-6}$ &  72 & 25   & 8.9$\times$10$^{-5}$ & 0.007 &   4.4 &   4.4 &   5.4 \\
 HR 333         &    41   & 2.5 & 110$\pm$15  & 9.8$\times$10$^{-5}$ &  34 &  6.4 & 1.1$\times$10$^{-4}$ & 0.02  &   7.8 &  33   &  31   \\
 49 Cet         &    19   & 2.2 & 118$\pm$6   & 3.0$\times$10$^{-4}$ &  22 &  3.5 & 7.7$\times$10$^{-5}$ & 0.01  &   7.6 & 100   &  19   \\
 HR 506         &     1.7 & 1.2 &  70$\pm$110 & 1.5$\times$10$^{-4}$ &  21 &  0.6 & 5.1$\times$10$^{-6}$ & 0.006 &   9.9 &  18   &  34   \\
 $\gamma$ Tri   &    54   & 2.6 & 120$\pm$15  & 2.2$\times$10$^{-5}$ &  31 &  8.1 & 2.7$\times$10$^{-5}$ & 0.003 &   8.4 &  56   &  24   \\
 $\tau^{3}$ Eri &    13   & 2.0 & 150$\pm$20  & 2.1$\times$10$^{-5}$ &  12 &  2.5 & 1.2$\times$10$^{-6}$ &0.0002 &  15   &  24   &  17   \\
 HR 1082        &    11   & 1.9 &  70$\pm$70  & 4.2$\times$10$^{-4}$ &  52 &  2.1 & 3.8$\times$10$^{-4}$ & 0.07  &   2.9 &  20   &  48   \\
 HR 1570        &    22   & 2.2 &  90$\pm$60  & 2.9$\times$10$^{-5}$ &  42 &  3.9 & 3.2$\times$10$^{-5}$ & 0.005 &  25   &  15   &  46   \\
 $\zeta$ Lep    &    30   & 2.3 & 191$\pm$3   & 6.5$\times$10$^{-5}$ &  10 &  5.0 & 5.6$\times$10$^{-6}$ & 0.0008&  36   & 250   &  83   \\
 HD 53143       &     0.48& 0.8 & 120$\pm$60  & 1.0$\times$10$^{-4}$ &   4 &  0.2 & 6.1$\times$10$^{-8}$ &0.00003&   2.4 &   6.4 &  23   \\
 HR 3314        &    43   & 2.5 & 130$\pm$15  & 2.5$\times$10$^{-5}$ &  27 &  6.8 & 1.9$\times$10$^{-5}$ & 0.002 &  12   &  46.5 &  33   \\
 HD 95086       &     7.5 & 1.7 &  80$\pm$30  & 6.4$\times$10$^{-4}$ &  30 &  1.8 & 1.7$\times$10$^{-4}$ & 0.03  &   5.1 &  23   &  46   \\
 $\beta$ UMa    &    80   & 2.7 & 110$\pm$30  & 1.1$\times$10$^{-5}$ &  53 & 12   & 5.3$\times$10$^{-5}$ & 0.005 &  30   &  72   &  72   \\
 $\beta$ Leo    &    14   & 2.0 & 120$\pm$15  & 2.7$\times$10$^{-5}$ &  19 &  2.7 & 4.2$\times$10$^{-6}$ & 0.0007&   2.3 & 190   &  38   \\
 HD 110058      &    10   &  ?  & 112$\pm$7   & 1.0$\times$10$^{-3}$ &  20 &   ?  &         ?            &   ?   &   4.3 &  57   & 170   \\
 $\lambda$ Boo  &    21   & 2.2 & 120$\pm$20  & 3.2$\times$10$^{-5}$ &  23 &  3.8 & 9.7$\times$10$^{-6}$ & 0.001 &  11   &  45   &  23   \\
 $\alpha$ CrB   &    85   & 2.7 & 139$\pm$7   & 1.3$\times$10$^{-5}$ &  33 & 12   & 2.7$\times$10$^{-5}$ & 0.002 &   3.5 & 140   &  54   \\
 HD 139664      &     3.6 & 2.7 & 100$\pm$60  & 3.8$\times$10$^{-5}$ &  15 &  1.0 & 1.2$\times$10$^{-6}$ & 0.0003&   7.0 &  23   &   7.5 \\
 HD 146897      &     4.6 & 1.5 & 100$\pm$8   & 5.4$\times$10$^{-3}$ &  17 &  1.2 & 2.9$\times$10$^{-4}$ & 0.06  &   4.2 &  70   & 240   \\ 
 HR 6297        &    11   & 1.8 & 110$\pm$40  & 5.1$\times$10$^{-5}$ &  21 &  2.3 & 8.1$\times$10$^{-6}$ & 0.001 &   8.4 &  10   &  15   \\
 HR 6486        &     6.8 & 1.7 & 120$\pm$20  & 6.7$\times$10$^{-5}$ &  14 &  1.6 & 3.3$\times$10$^{-6}$ & 0.0006&   4   &  38   &   6.9 \\
 HR 6532        &    93   & 3.0 & 120$\pm$60  & 2.9$\times$10$^{-5}$ &  47 & 12   & 1.2$\times$10$^{-4}$ & 0.01  &   3.1 &   7.2 &   3.6 \\
 78 Her         &    51   & 2.6 & 140$\pm$160 & 1.6$\times$10$^{-5}$ &  23 &19    & 2.5$\times$10$^{-5}$ & 0.002 &   0.5 &   6.5 &   0.7 \\
 $\gamma$ Oph   &    39   & 2.5 & 124$\pm$9   & 5.2$\times$10$^{-5}$ &  27 &  6.1 & 3.8$\times$10$^{-5}$ & 0.005 &  29   & 130   & 120   \\
 HR 6670        &     4.1 & 1.5 & 110$\pm$120 & 2.4$\times$10$^{-5}$ &  13 &  1.1 & 7.4$\times$10$^{-7}$ &0.0002 &   7.3 &   2.4 &   9.6 \\
 HD 181327      &     2.5 & 1.9 &  81$\pm$7   & 3.1$\times$10$^{-3}$ &  20 &  0.9 & 1.7$\times$10$^{-4}$ & 0.04  &  21   & 104   &1500   \\
 HD 191089      &     3.1 & 1.8 &  99$\pm$9   & 1.5$\times$10$^{-3}$ &  14 &  0.9 & 4.4$\times$10$^{-5}$ & 0.01  &   3.6 &  75   & 220   \\
 HR 8799        &     6.1 & 1.6 & 150$\pm$30  & 4.9$\times$10$^{-5}$ &   8 &  1.5 & 7.2$\times$10$^{-7}$ & 0.0002&   4.7 &  13   &   3.1 \\
\enddata
\end{deluxetable}
\clearpage
\pagestyle{plaintop}
\begin{deluxetable}{lc|cc|cc|c|c}
\small
\rotate
\tablecaption{Dust Removal Mechanisms}
\tablehead{
\omit &
\omit &
\multicolumn{2}{c}{PR Drag} &
\multicolumn{2}{c}{CPR Drag} &
\colhead{Collision} &
\colhead{Sublimation} \\
\colhead{Name} &
\colhead{$\frac{\dot{M}_{wind} c^{2}}{L_{*}}$} &
\colhead{$t_{PR}$} &
\colhead{$M_{PB}$} &
\colhead{$t_{PR+wind}$} &
\colhead{$M_{PB}$} &
\colhead{$t_{coll}$} &
\colhead{$t_{subl}$} \\
\omit &
\omit &
\colhead{(Myr)} &
\colhead{($M_{\earth}$)} &
\colhead{(Myr)} &
\colhead{($M_{\earth}$)} &
\colhead{(Myr)} &
\colhead{(Myr)} \\
}
\tablewidth{0pt}
\tablecolumns{8}
\startdata
 $\lambda$ Cas  & 1.4 &  0.88  & 0.012  & 0.36     & 0.0034 & 1.9    & 7.3$\times$10$^{-4}$ \\ 
 HR 333         & ... &  0.31  & 0.040  &   ...    &  ...   & 0.034  & 7.8$\times$10$^{-3}$ \\ 
 HR 506         &82   &  0.24  & 0.0063 & 0.003    & 0.13   & 0.020  & 1.1$\times$10$^{8}$  \\ 
 49 Cet         & ... &  0.15  & 0.058  &   ...    &  ...   & 0.0073 & 9.3$\times$10$^{-5}$ \\ 
 $\gamma$ Tri   & ... &  0.25  & 0.019  &   ...    &  ...   & 0.14   & 3.0$\times$10$^{-5}$ \\ 
 $\tau^{3}$ Eri & ... &  0.049 & 0.012  &   ...    &  ...   & 0.057  & 3.1$\times$10$^{-9}$\\ 
 HR 1082        & ... &  0.93  & 0.061  &   ...    &  ...   & 0.025  & 1.3$\times$10$^{10}$ \\ 
 HR 1570        & ... &  0.54  & 0.013  &   ...    &  ...   & 0.22   & 2.0$\times$10$^{2}$  \\ 
 $\zeta$ Lep    & ... &  0.031 & 0.032  &  ...     &  ...   & 0.011  & 3.5$\times$10$^{-12}$\\ 
 HD 53143       & ?   &  0.014 & 0.0013 & ?        &   ?    & 0.0036 & 1.5$\times$10$^{-5}$ \\ 
 HR 3314        & ... &  0.20  & 0.018  &   ...    &  ...   & 0.11   & 3.9$\times$10$^{-6}$ \\ 
 HD 95086       & ... &  0.37  & 0.0071 &   ...    &  ...   & 0.0081 & 7.3$\times$10$^{3}$  \\ 
 $\beta$ UMa    & ... &  0.69  & 0.026  &   ...    &  ...   & 0.7    & 2.7$\times$10$^{-2}$ \\ 
 $\beta$ Leo    & ... &  0.12  & 0.0049 &   ...    &  ...   & 0.080  & 3.8$\times$10$^{-5}$ \\ 
 HD 110058      & ... &   ...  & 0.015  &   ...    &  ...   &  ...   & ... \\
 $\lambda$ Boo  & ... &  0.16  & 0.017  &   ...    &  ...   & 0.085  & 2.5$\times$10$^{-5}$ \\ 
 $\alpha$ CrB   & ... &  0.26  & 0.031  &   ...    &  ...   & 0.28   & 3.4$\times$10$^{-7}$ \\ 
 HD 139664      & ?   &  0.10  & 0.0059 & ?        & ?      & 0.058  & 1.7$\times$10$^{-1}$ \\ 
 HD 146897      & ?   &  0.13  & 0.011  & ?        & ?      & 0.00042& 1.4$\times$10$^{-1}$ \\ 
 HR 6297        & ... &  0.17  & 0.021  &   ...    &  ...   & 0.027  & 2.0$\times$10$^{-3}$ \\ 
 HR 6486        & ... &  0.08  & 0.0099 &   ...    &  ...   & 0.026  & 1.2$\times$10$^{-5}$ \\ 
 HR 6532        & 1.2 &  0.48  & 0.050  & 0.17     & 0.015  & 0.2    & 2.2$\times$10$^{-4}$ \\ 
 78 Her         & ... &  0.34  & 0.0036 &   ...    &  ...   & 0.29   & 4.3$\times$10$^{-7}$ \\ 
 $\gamma$ Oph   & ... &  0.20  & 0.034  &   ...    &  ...   & 0.052  & 2.3$\times$10$^{-5}$ \\ 
 HR 6670        & 190 &  0.081 & 0.015  & 0.00043  & 0.69   & 0.068  & 1.9$\times$10$^{-3}$ \\ 
 HD 181327      & ?   &  0.24  & 0.99   & ?        &  ?     & 0.0013 & 6.0$\times$10$^{4}$  \\ 
 HD 191089      & ?   &  0.11  & 0.65   & ?        &  ?     & 0.0014 & 2.7$\times$10$^{-1}$ \\ 
 HR 8799        & ... &  0.027 & 0.016  &   ...    &  ...   & 0.014  & 1.3$\times$10$^{-9}$ \\ 
\enddata
\end{deluxetable}
\begin{deluxetable}{lccccc}
\small
\tablecaption{X-ray and Stellar Wind Properties of FGKM Stars}
\tablehead{
\colhead{Name} &
\colhead{Offset} &
\colhead{\emph{ROSAT}} &
\colhead{HR1} &
\colhead{$F_{X}$} &
\colhead{$\dot{M}_{wind}$} \\
\omit &
\colhead{($\arcsec$)} &
\colhead{(cts s$^{-1}$)} &
\omit &
\colhead{(erg s$^{-1}$ cm$^{-2}$)} &
\colhead{($\dot{M}_{\sun}$)} \\
}
\tablewidth{0pt}
\tablecolumns{6}
\startdata
107 PSC      & 0.13 & 0.027$\pm$0.003 & -0.92$\pm$0.04 &    5.1$\times$10$^{3}$ &        3.0 \\
HR 506       & 0.46 & 0.092           & -0.63$\pm$0.03 &    1.7$\times$10$^{5}$ &      290   \\
HD 16157     & 0.10 & 3.52$\pm$0.02   &  0.11$\pm$0.01 &                        &            \\
$\tau^1$ Eri & 0.04 & 0.72$\pm$0.02   & -0.25$\pm$0.03 &    1.0$\times$10$^{6}$ &       ?    \\
$\alpha$ For & 0.20 & 2.8$\pm$0.2     & -0.08$\pm$0.05 &    2.3$\times$10$^{6}$ &       ?    \\
$\gamma$ Dor & 0.11 & 0.19$\pm$0.04   & -0.09$\pm$0.18 &    6.7$\times$10$^{5}$ &     2100   \\
HR 1686      & 0.18 & 0.020$\pm$0.007 & -1.0$\pm$0.4   &    1.5$\times$10$^{4}$ &       26   \\
HR 2483      & ...  & $<$0.05         & ...            & $<$1.5$\times$10$^{5}$ &   $<$260   \\
HD 53143     & 0.48 & 0.19$\pm$0.01   & -0.37$\pm$0.05 &    8.8$\times$10$^{5}$ &       ?    \\
HR 3220      & 0.14 & 0.35$\pm$0.03   & -0.18$\pm$0.09 &    7.1$\times$10$^{5}$ &     4100   \\
FI Vir       & 0.05 & 0.053           & -0.50$\pm$0.06 &                        &            \\
HR 3862      & 0.07 & 0.002$\pm$0.0007& ...            & $<$2.9$\times$10$^{3}$ &     $<$2.1 \\
$\eta$ Crv   & 0.02 & 0.21$\pm$0.03   & -0.48$\pm$0.13 &    3.4$\times$10$^{5}$ &     1100   \\
HD 113766    & ...  & $<$0.05         & ...            & $<$7.6$\times$10$^{6}$ &       ?    \\
HR 5236      & ...  & $<$0.05         & ...            & $<$7.2$\times$10$^{5}$ &  $<$2200   \\
$\sigma$ Boo & 0.11 & 0.17$\pm$0.02   & -0.32$\pm$0.11 &    2.5$\times$10$^{5}$ &      650   \\
HD 139664    & 0.11 & 0.52$\pm$0.04   & -0.28$\pm$0.06 &    1.2$\times$10$^{6}$ &       ?    \\
HR 5933      & 0.35 & 0.04$\pm$0.02   & -0.90$\pm$0.14 &    1.2$\times$10$^{4}$ &       14   \\
HD 146897    & ...  & $<$0.05         & ...            & $<$5.6$\times$10$^{6}$ &       ?    \\
HR 6670      & 0.03 & 0.10$\pm$0.02   & -0.50$\pm$0.17 &    4.4$\times$10$^{5}$ &     1700   \\
HD 181327    & ...  & $<$0.05         & ...            & $<$1.3$\times$10$^{6}$ &       ?    \\
HD 191089    & 0.06 & 0.02$\pm$0.02   & -0.30$\pm$0.22 &    1.4$\times$10$^{6}$ &       ?    \\
HD 221354    & ...  & $<$0.05         & ...            & $<$3.7$\times$10$^{4}$ &   $<$160   \\
\enddata
\end{deluxetable}
\begin{deluxetable}{lcccccc}
\small
\tablecaption{Binary Properties}
\tablehead{
\colhead{Name} &
\colhead{Primary \& Secondary} &
\colhead{Separation} &
\colhead{Separation} &
\colhead{Period} &
\colhead{Notes$^{\dagger}$} &
\colhead{References} \\
\omit &
\colhead{Spectral Types} &
\colhead{($\arcsec$)} &
\colhead{(AU)} &
\colhead{(years)} &
\omit &
\omit \\
}
\tablewidth{0pt}
\tablecolumns{6}
\startdata
 $\lambda$ Cas &    B7V \& B8V   &    0.38 &   41 & 515  & VB  & 2 \\
 HR 1570       &    A0V \& ?     &   40    & 1500 &      &     & 8 \\
 HR 1686       &    F6V \& ?     &   30    &  630 &      &     & 8 \\
 $\kappa$ Lep  &    B7V \& B9V   &    2.2  &  370 &      &     & 3,6 \\
 32 Ori        &    B5IV \& B7V  &    1.3  &  120 & 590  & VB  & 1 \\
 HR 2124       &    A2V \& ?     &    0.1  &    5 &      &     & 5 \\
 $\psi^5$ Aur  &    G0V \& M0V   &   31    &  520 &      &     & 8 \\
 G Cen         &    B3Vn \& ?    &   23    & 2800 &      &     & 8 \\
 HD 113766     &   F3/F5V \& ?   &    1.3  &  170 &      &     & 3 \\
 HR 6532       & B9.5V \& $<$A9V &         &      &0.019 & SB  & 4 \\
 HR 7329       &  A0Vn \& M7/8V  &    4    &  200 &1700  & CPM & 7 \\
\enddata
\tablenotetext{\dagger}{CPM = Common Proper Motion, SB = Spectroscopic Binary, 
  VB = Visual Binary}
\tablerefs{(1) Abt \& Cardona 1984; (2) Abt \& Boonyarak 2004; 
  (3) Fabricius \& Makarov 2000; (4) Guthrie et al. 1986; 
  (5) Horsch et al. 2002; (6) Lindroos 1985; (7) Lowrance et al 2000; 
  (8) Simbad}
\end{deluxetable}
\begin{deluxetable}{lccccc}
\small
\tablecaption{Emission Line and Gas Mass 3$\sigma$ Upper Limits}
\tablehead{
\omit &
\omit &
\omit &
\omit &
\colhead{50 K} &
\colhead{100 K} \\
\colhead{Name} &
\colhead{H$_{2}$ S(1)} &
\colhead{\ion{S}{1}} &
\colhead{H$_{2}$ S(0)} &
\colhead{$M_{H_{2}}$} &
\colhead{$M_{H_{2}}$} \\
\omit &
\colhead{(erg s$^{-1}$ cm$^{-2}$)} &
\colhead{(erg s$^{-1}$ cm$^{-2}$)} &
\colhead{(erg s$^{-1}$ cm$^{-2}$)} &
\colhead{$(M_{\earth})$} &
\colhead{$(M_{\earth})$} \\
}
\tablewidth{0pt}
\tablecolumns{6}
\startdata
49 Cet       & $<$8.6$\times$10$^{-15}$ & $<$1.0$\times$10$^{-14}$ & $<$7.4$\times$10$^{-15}$ & $<$6000 & $<$80 \\
$\tau$ Cet   & $<$2.7$\times$10$^{-13}$ & $<$3.2$\times$10$^{-14}$ & $<$1.9$\times$10$^{-14}$ &   $<$70 &  $<$1 \\
$\tau^3$ Eri & $<$5.0$\times$10$^{-15}$ & $<$1.5$\times$10$^{-14}$ & $<$5.8$\times$10$^{-15}$ &  $<$900 & $<$10 \\
$\alpha$ For & $<$4.3$\times$10$^{-14}$ & $<$1.4$\times$10$^{-14}$ & $<$1.3$\times$10$^{-14}$ &  $<$600 &  $<$7 \\
$\zeta$ Lep  & $<$2.1$\times$10$^{-14}$ & $<$1.8$\times$10$^{-14}$ & $<$1.0$\times$10$^{-14}$ & $<$1000 & $<$10 \\
HR 2124      & $<$2.3$\times$10$^{-14}$ & $<$7.7$\times$10$^{-15}$ & $<$1.1$\times$10$^{-14}$ & $<$5000 & $<$70 \\
$\delta$ Vel & $<$5.0$\times$10$^{-14}$ & $<$1.2$\times$10$^{-14}$ & $<$4.2$\times$10$^{-15}$ &  $<$500 &  $<$7 \\
$\beta$ UMa  & $<$1.9$\times$10$^{-14}$ & $<$7.3$\times$10$^{-15}$ & $<$3.8$\times$10$^{-15}$ &  $<$500 &  $<$6 \\
$\beta$ Leo  & $<$2.5$\times$10$^{-14}$ & $<$6.1$\times$10$^{-14}$ & $<$4.4$\times$10$^{-14}$ & $<$1000 & $<$10 \\
$\eta$ Crv   & $<$2.3$\times$10$^{-14}$ & $<$1.2$\times$10$^{-14}$ & $<$8.9$\times$10$^{-15}$ &  $<$600 &  $<$8 \\
HD 113766    & $<$1.1$\times$10$^{-13}$ & $<$2.1$\times$10$^{-14}$ & $<$1.0$\times$10$^{-14}$ &$<$40000 &$<$500 \\
$\gamma$ Tra & $<$2.7$\times$10$^{-14}$ & $<$1.4$\times$10$^{-14}$ & $<$9.3$\times$10$^{-15}$ & $<$7000 & $<$80 \\
$\alpha$ CrB & $<$3.0$\times$10$^{-14}$ & $<$1.3$\times$10$^{-14}$ & $<$6.5$\times$10$^{-15}$ &  $<$800 & $<$10 \\
41 Ser       & $<$3.7$\times$10$^{-14}$ & $<$1.3$\times$10$^{-14}$ & $<$8.0$\times$10$^{-15}$ &  $<$200 &  $<$3 \\
HR 6486      & $<$2.6$\times$10$^{-14}$ & $<$1.0$\times$10$^{-14}$ & $<$9.0$\times$10$^{-15}$ & $<$1000 & $<$20 \\
$\gamma$ Oph & $<$1.3$\times$10$^{-14}$ & $<$1.9$\times$10$^{-14}$ & $<$6.4$\times$10$^{-15}$ & $<$1000 & $<$20 \\
HR 7012      & $<$2.4$\times$10$^{-14}$ & $<$1.1$\times$10$^{-14}$ & $<$1.3$\times$10$^{-14}$ & $<$2400 & $<$30 \\
\enddata
\end{deluxetable}


\begin{figure}
\figurenum{1}
\epsscale{1}
\plotone{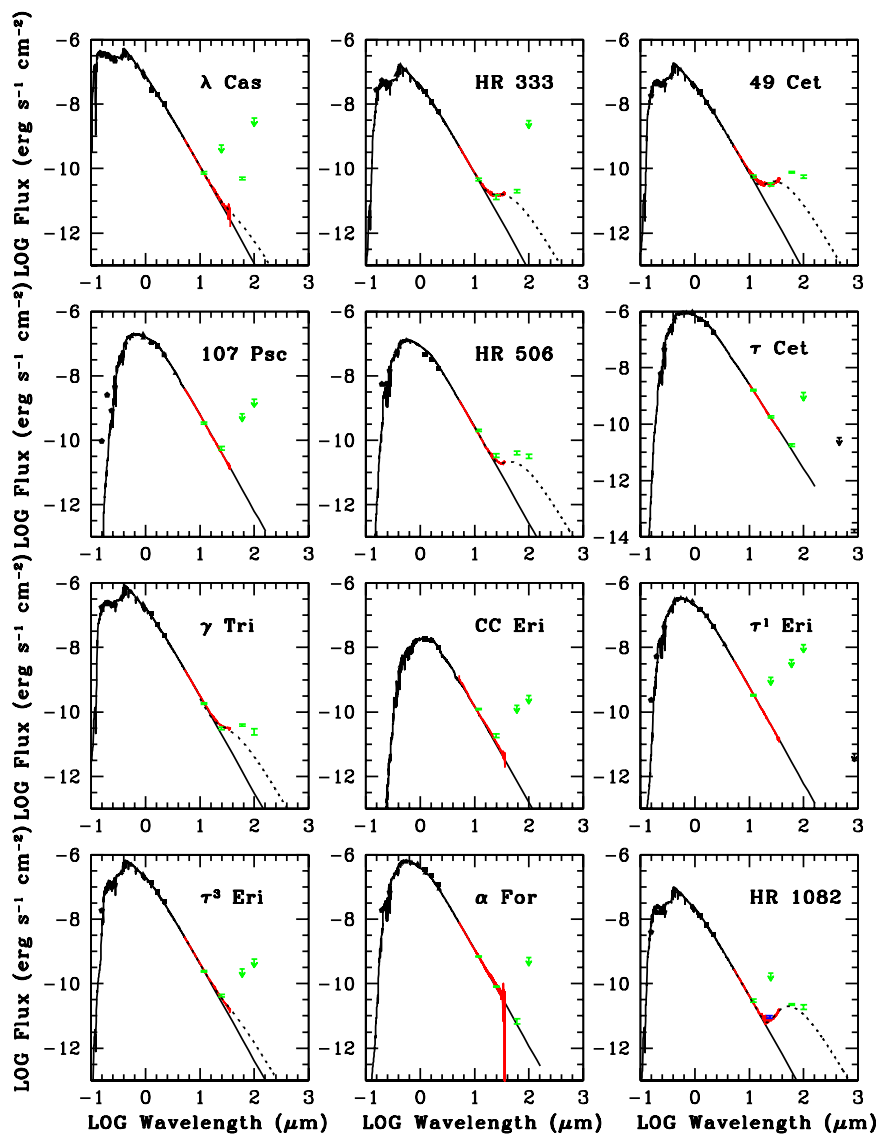}
\caption{Spectral Energy Distributions (SEDs) for all objects in our sample.  
TD1 fluxes (Thompson et al. 1978) are plotted as pentagons, General Catalogue
of Photometric Data mean $UBV$ or Johnson et al. (1966) fluxes are plotted as 
triangles, and 2MASS JHK fluxes (Cutri et al. 2003) are plotted as squares. 
Color-corrected \emph{IRAS}, MIPS, and submillimeter photometry, where 
available, are shown with green, blue, and black error bars and upper limit
symbols, respectively. Our \emph{Spitzer} IRS spectra, as reported here, are 
shown in red. Overlaid are the best fit 1993 Kurucz and Nextgen models for the 
stellar atmospheres.}
\end{figure}
\clearpage
\plotone{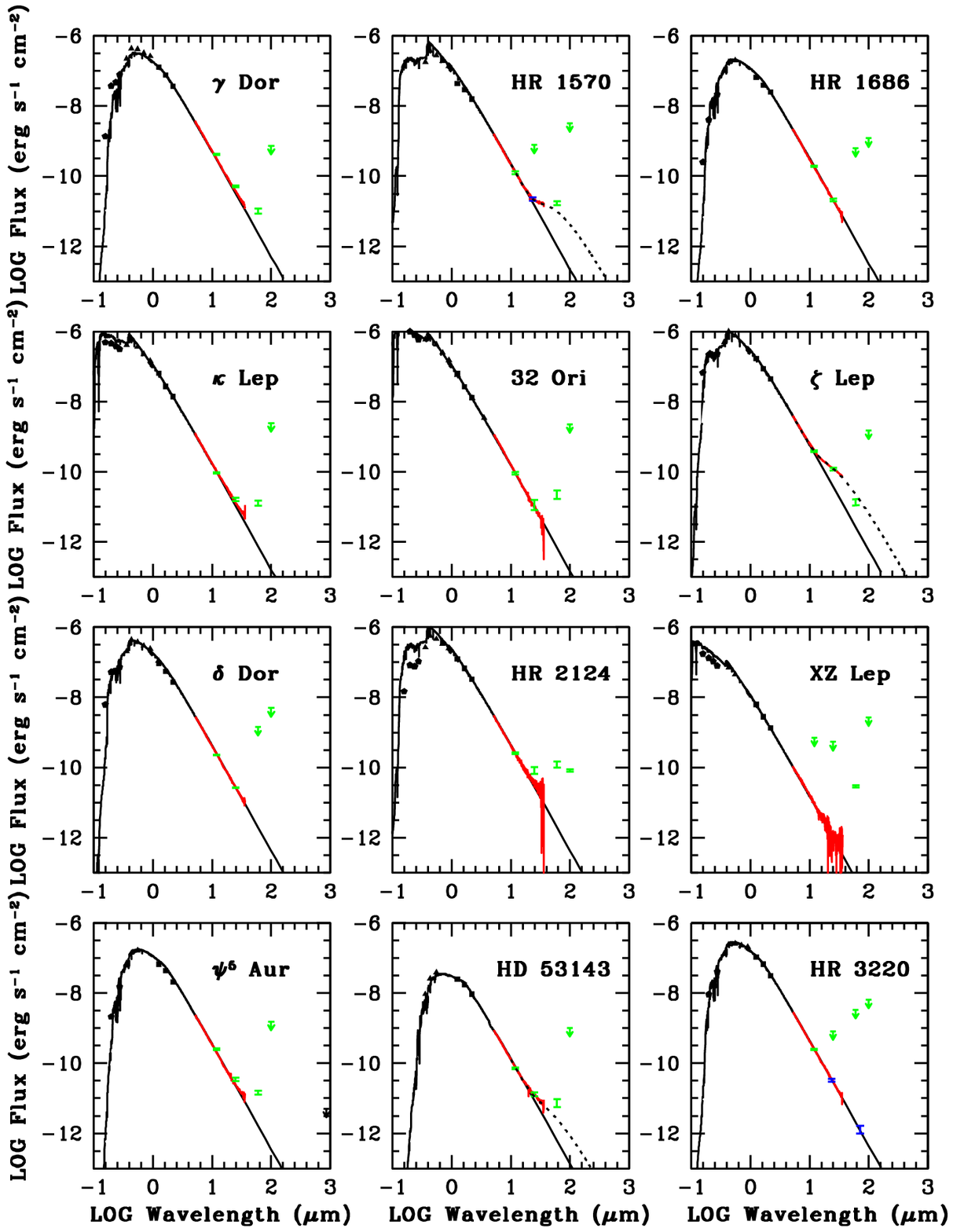}
\centerline{Fig. 1b. ---}
\clearpage
\plotone{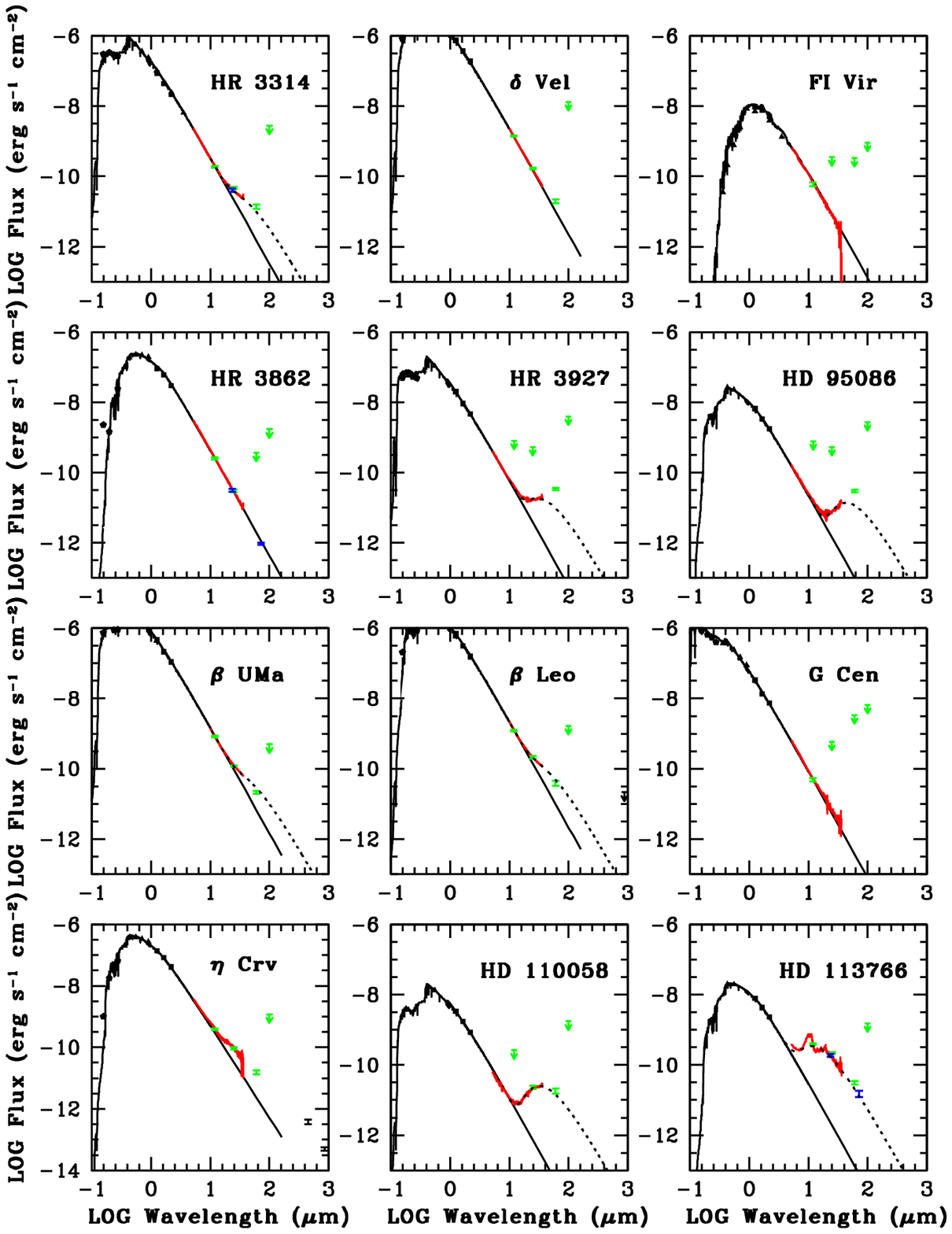}
\centerline{Fig. 1c. ---}
\clearpage
\plotone{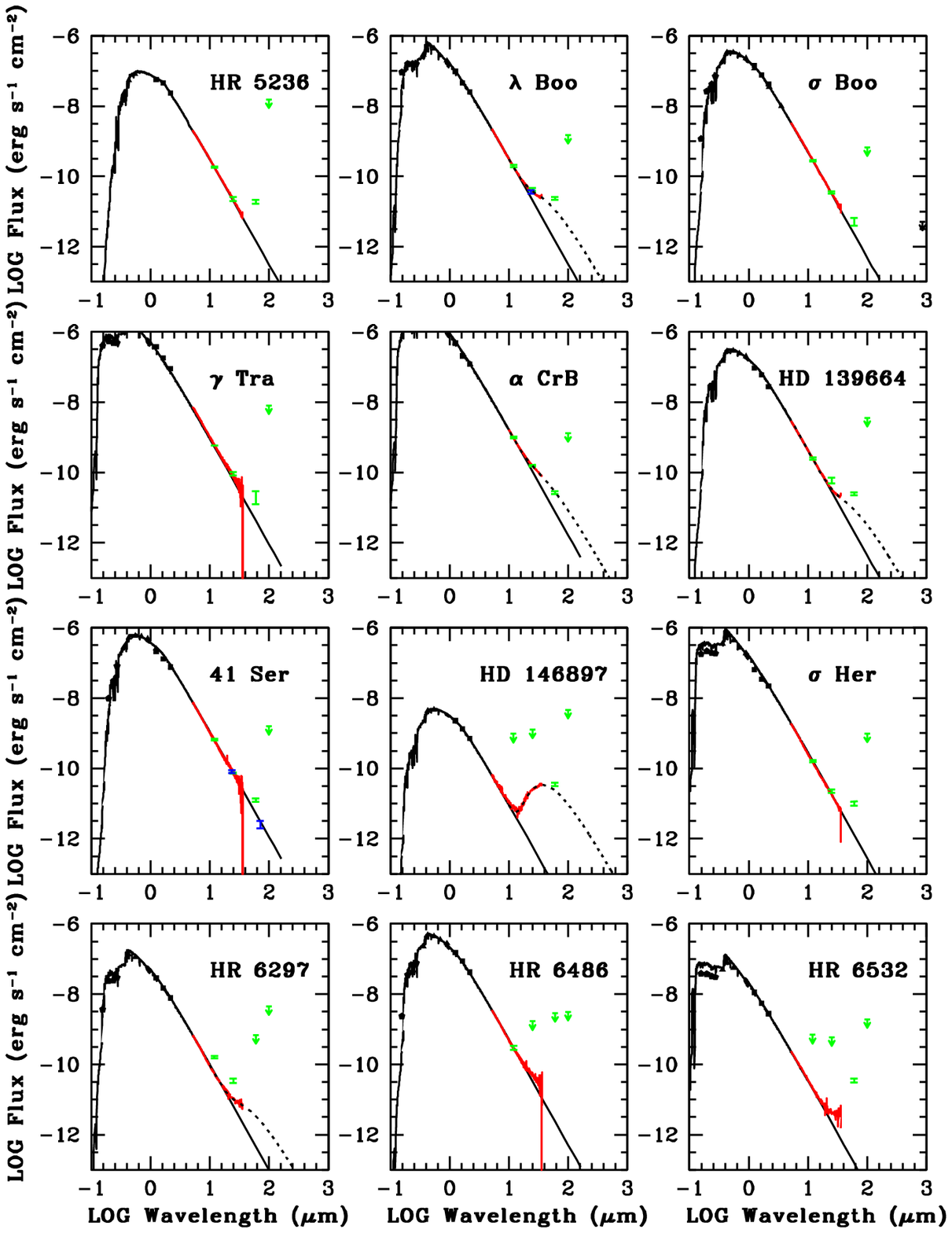}
\centerline{Fig. 1d. ---}
\clearpage
\plotone{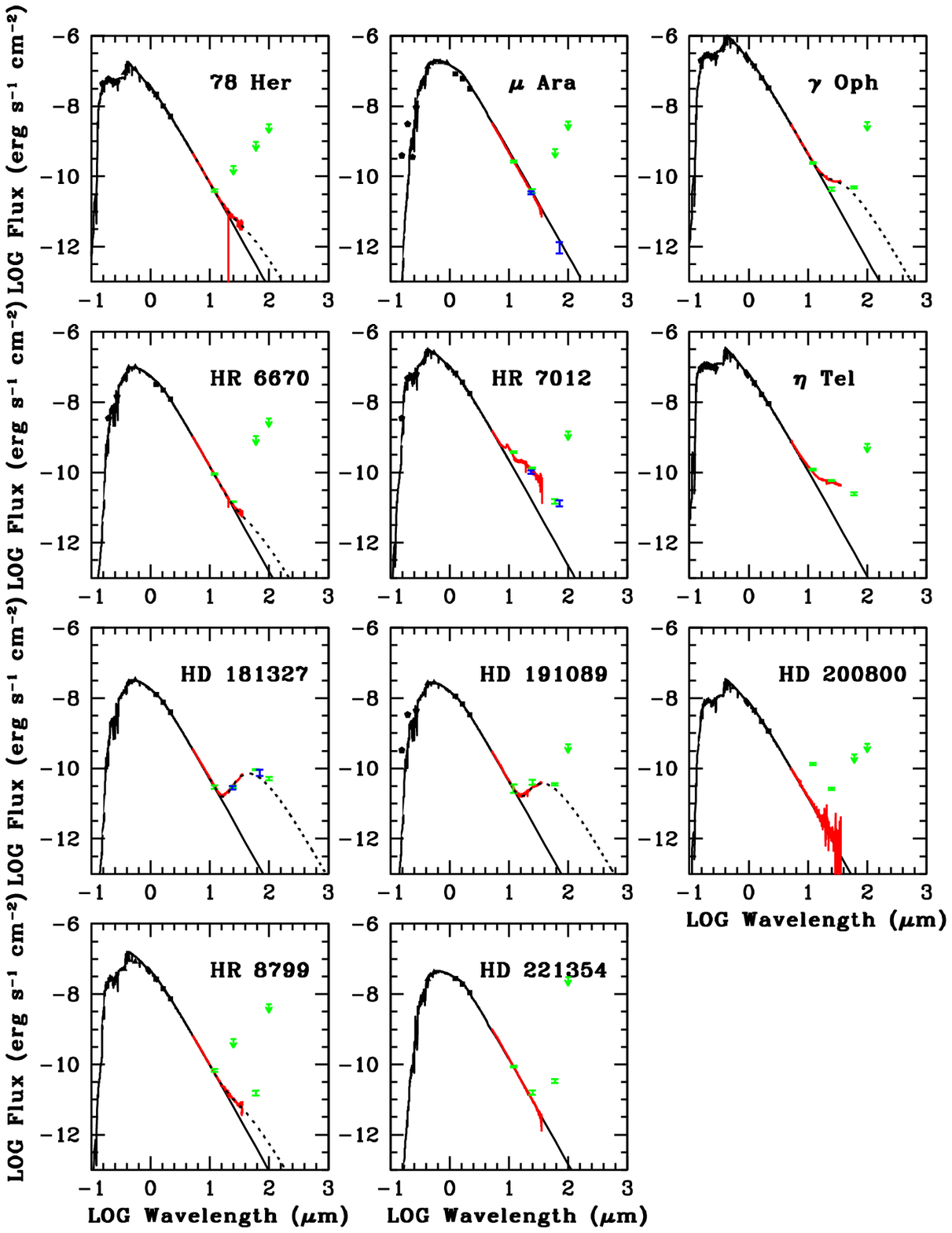}
\centerline{Fig. 1e. ---}
\clearpage
\begin{figure}
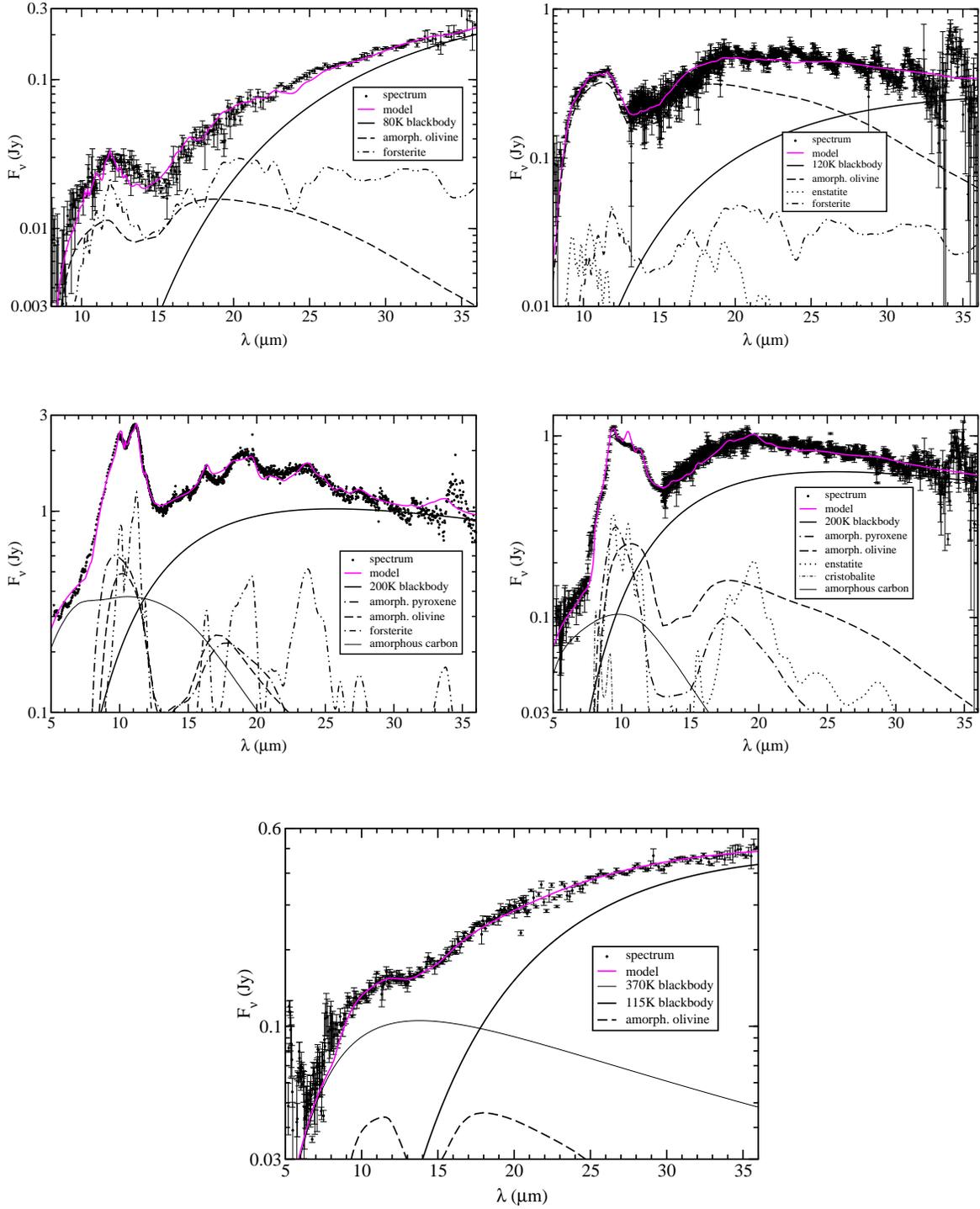

\figurenum{2}
\epsscale{1}
\plottwo{f2a.eps}{f2b.eps}
\vspace{9mm}
\plottwo{f2c.eps}{f2d.eps}
\vspace{10mm}
\epsscale{0.5}
\plotone{f2e.eps}
\caption{Photosphere-subtracted IRS spectra of (a) HR 3927, (b) $\eta$ Crv,
(c) HD 113766, (d) HR 7012, and (e) $\eta$ Tel. The solid magenta lines
show the final models for each disk system.}
\end{figure}

\begin{figure}
\figurenum{3a}
\epsscale{1}
\plotone{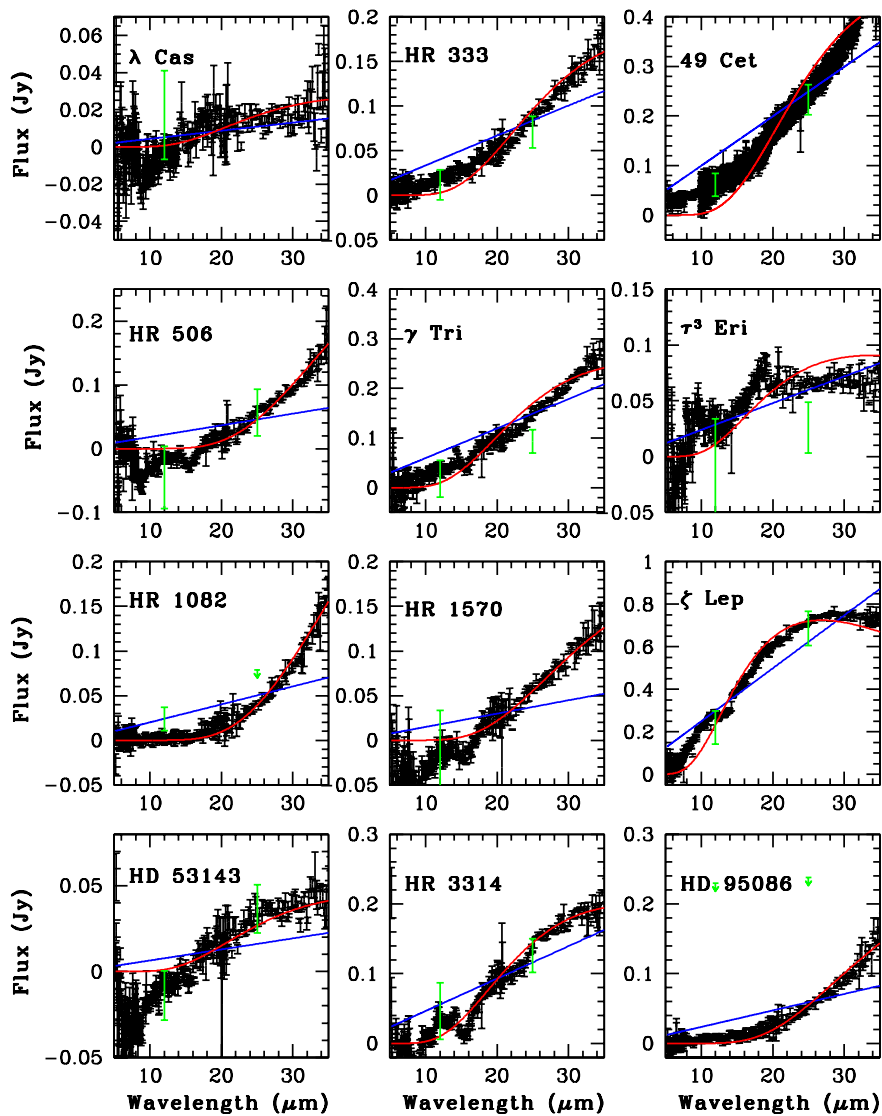}
\caption{Photosphere subtracted IRS spectra with $F_{\nu}$ plotted as a 
function of wavelength. The minimum $\chi^{2}$ fits for the single
temperature black body and continuous disk models are overplotted in
red and blue, respectively. \emph{IRAS} photosphere subtracted fluxes,
where available, are overplotted in green to highlight any discrepancies
with IRS data.}
\end{figure}
\clearpage
\plotone{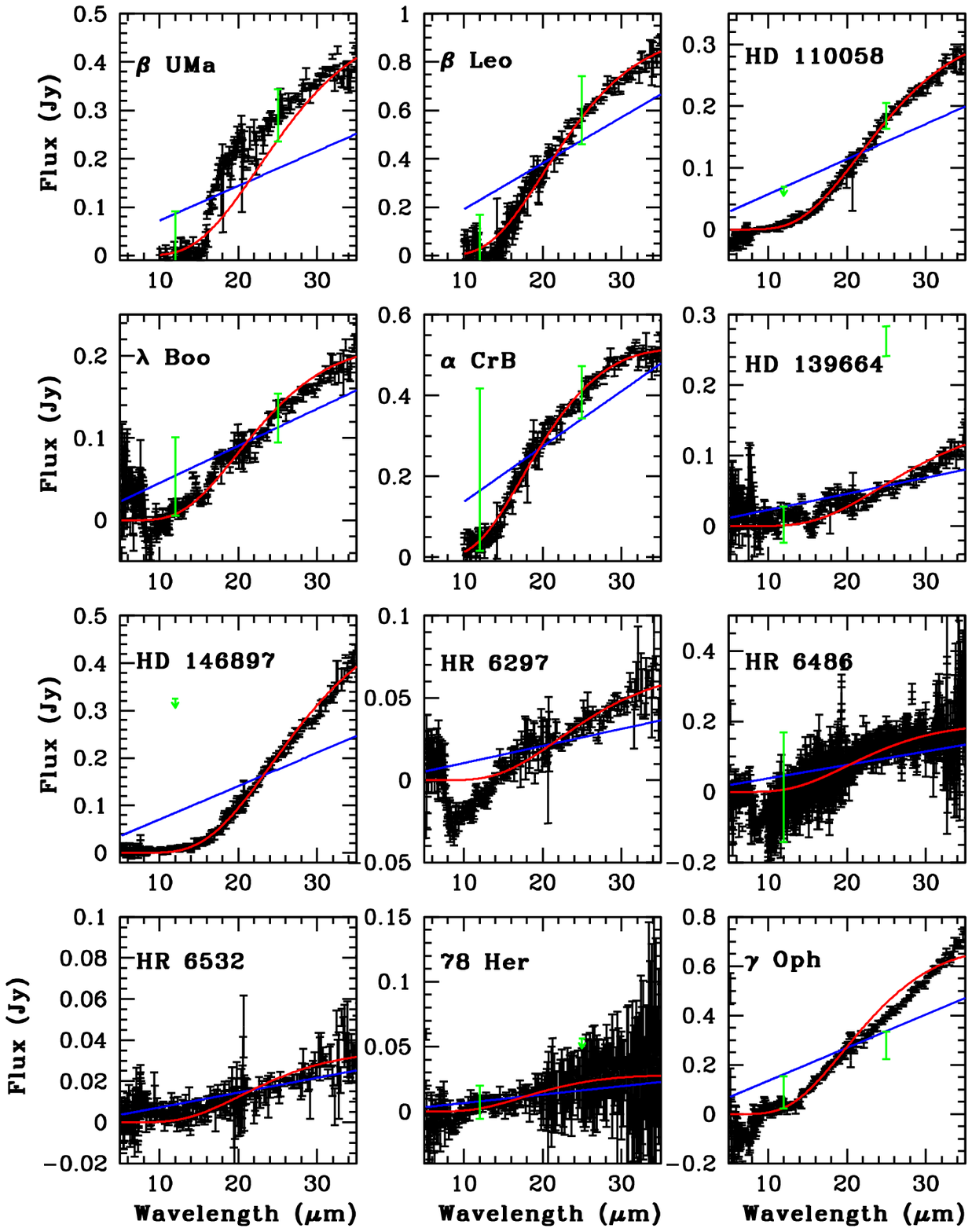}
\centerline{Fig. 3b ---}
\clearpage
\plotone{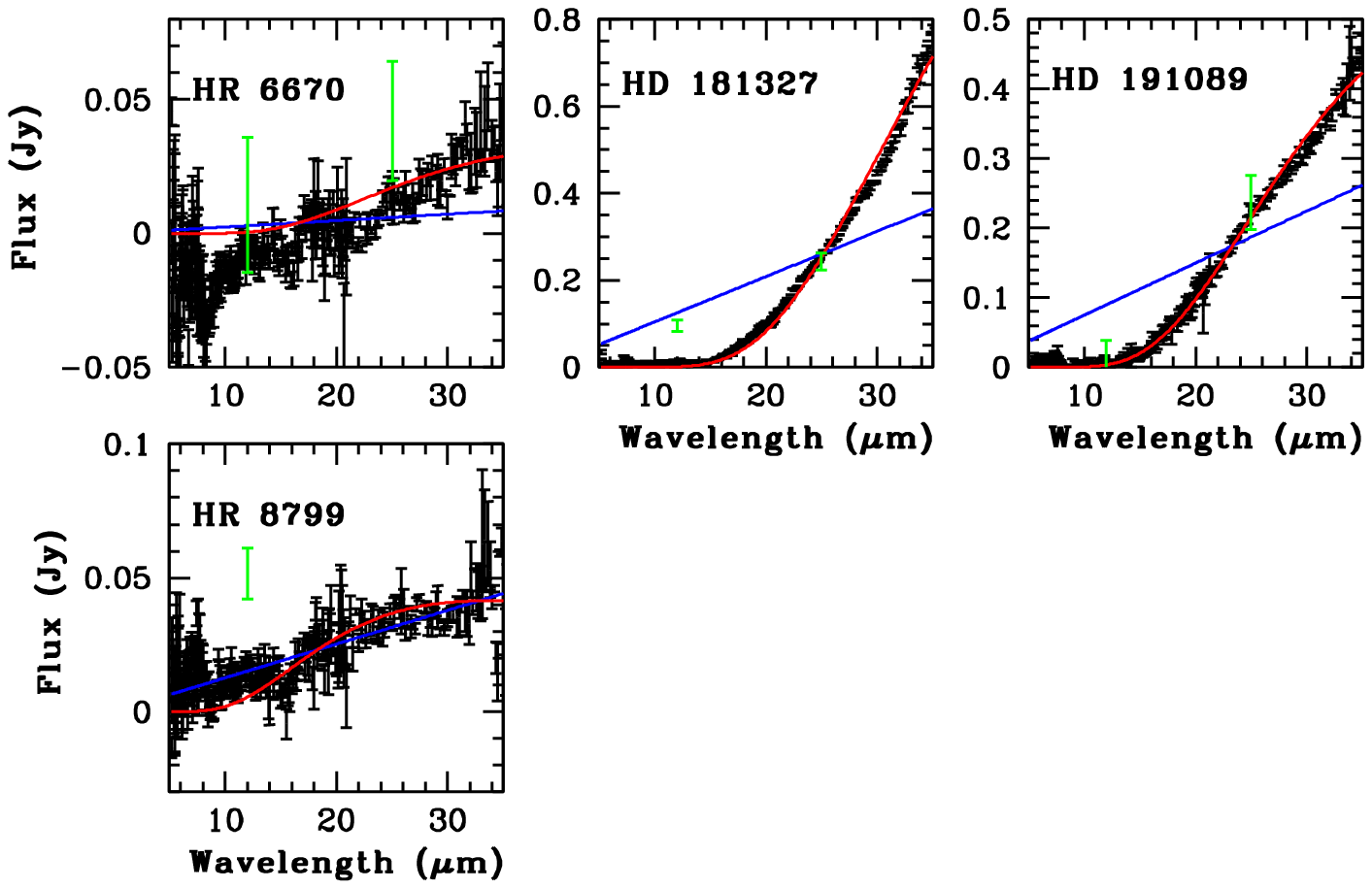}
\centerline{Fig. 3c ---}
\clearpage
\begin{figure}
\figurenum{4}
\epsscale{1}
\plotone{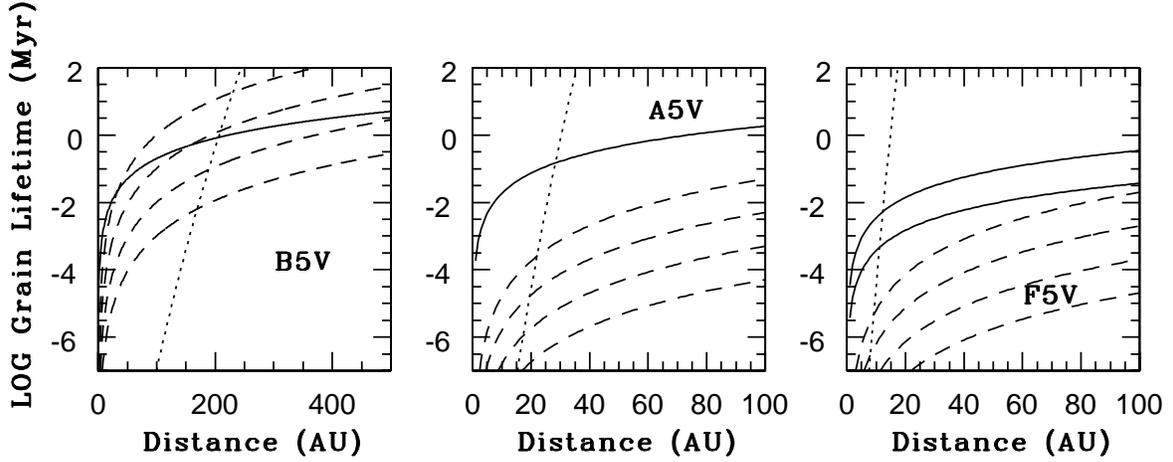}
\caption{(a) The grain lifetimes are plotted as a function of distance around a
B5V star. The Poynting-Robertson Drag/Stellar Wind drag lifetime is shown with 
a solid line; the sublimation lifetime is showed with a dotted line; and the 
collisional lifetime is shown with a dashed line, assuming $M_{submm}$ = 
0.001, 0.01, 0.1, and 1 $M_{\earth}$ (from top to bottom). (b) same as (a) 
for A5V star. No stellar wind drag is assumed. (c) same as (a) for a F5V star.
The Stellar Wind drag lifetime is shown with a solid line, assuming that
${\dot M_{wind}}$ = 100, 1000 ${\dot M_{\sun}}$ (from top to bottom).}
\end{figure}

\begin{figure}
\figurenum{5}
\plotone{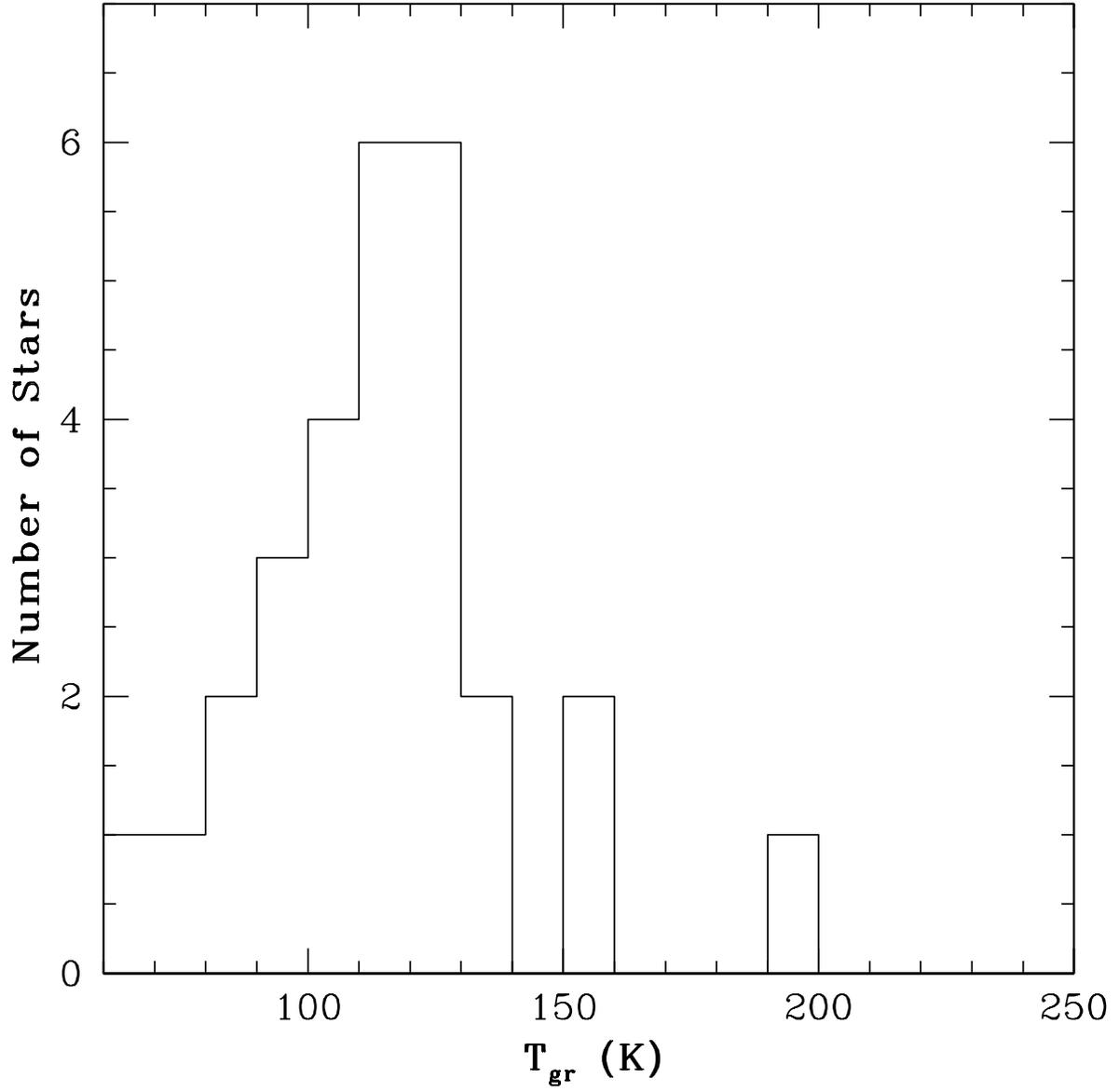}
\caption{Histogram showing the distribution of inferred black body grain 
temperatures for objects whose IRS spectra are well-fit by a single temperature
black body.}
\end{figure}

\begin{figure}
\figurenum{6}
\plottwo{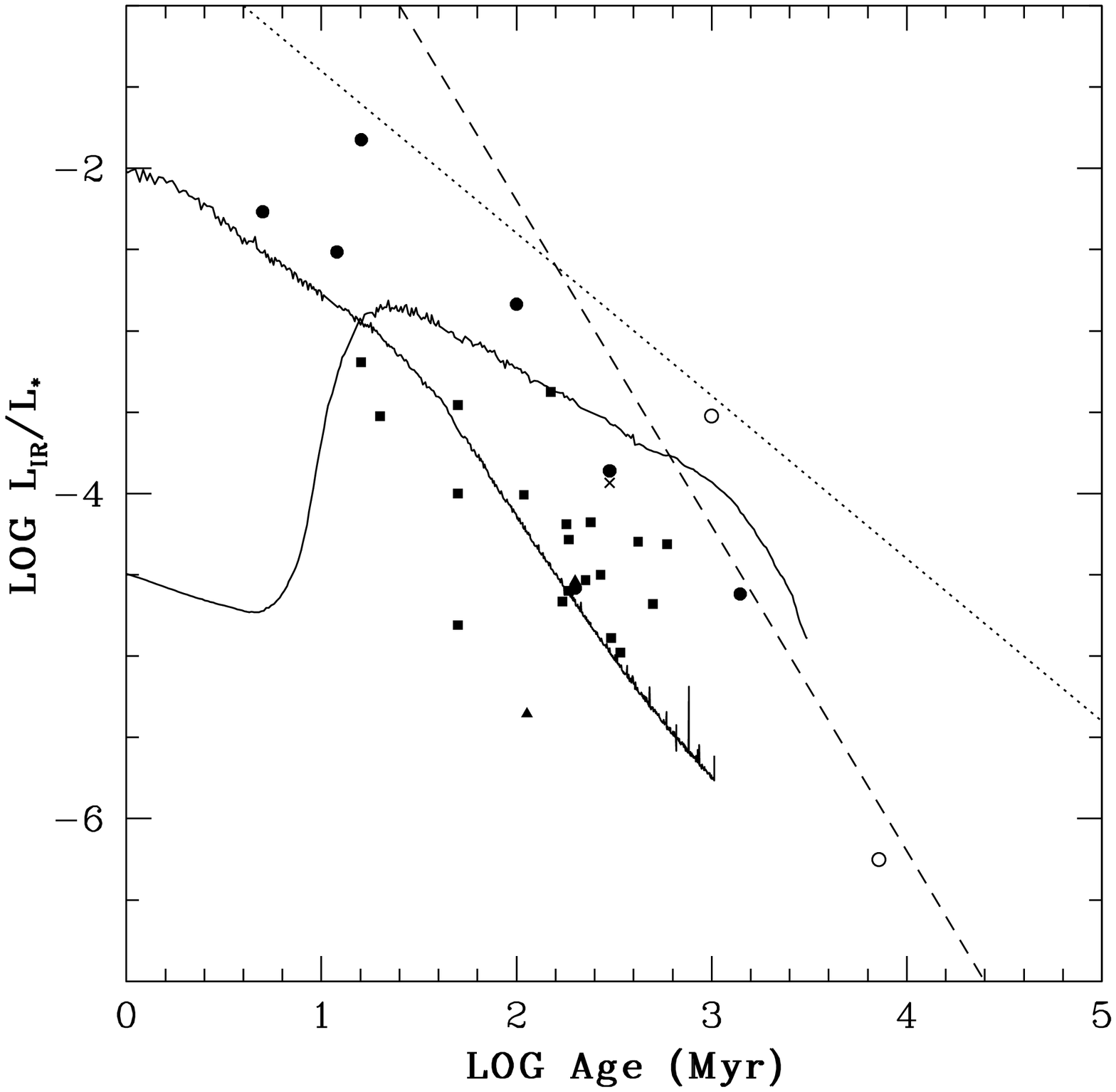}{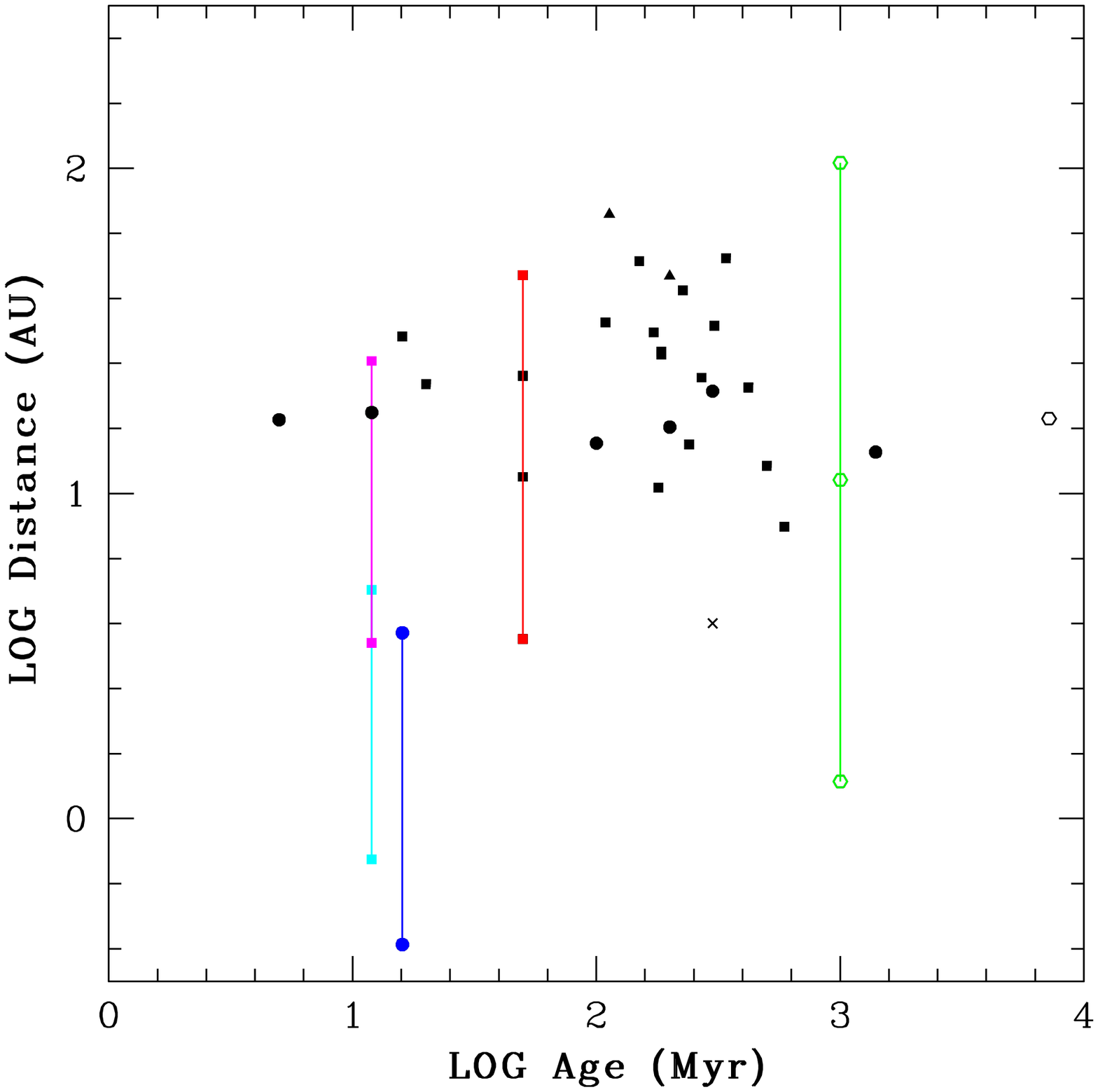}
\caption{(a) Inferred fractional luminosity, $L_{IR}/L_{*}$ plotted as a 
function of age. The solid lines show the Kenyon \& Bromley (2005) model for 
the evolution of $L_{IR}/L_{*}$ for dust at 3 - 20 AU (left) and at 30 - 100 AU
(right) around a main sequence A-type star. The dotted line shows the slope 
expected if the dust mass declines inversely with age as expected if collisions
are the dominant grain destruction mechanism; HD 113766 and $\eta$ Crv possess 
extremely high values of $L_{IR}/L_{*}$ for their ages and define the fitting 
coefficient for the $1/t$ trend line. The dashed line shows the slope expected 
if Poynting-Robertson and corpuscular stellar wind drag are the dominant
grain destruction mechanism; this trend line accounts for all of the stars
in our sample except for $\eta$ Crv. (b) Black body grain distance plotted as 
a function of stellar age. Stars with proposed multiple debris belts are 
shown in color with a line connecting multiple points. HR 3927, $\eta$ Crv,
HD 113766, HR 7012, and $\eta$ Tel are shown in red, green, blue, cyan, and
magenta, respectively. In both plots, B-type stars are shown with filled 
triangles; A-type stars are shown with filled squares; F-type stars are shown 
with filled circles; the K-type star is shown with a cross.  The addition of 
submillimeter data to infer $L_{IR}/L_{*}$ and $D$ for $\eta$ Crv and $\tau$ 
Cet (shown as open circles) does not appear to elucidate either the relation 
between fractional infrared luminosity and age or grain distance and age.}
\end{figure}

\begin{figure}
\figurenum{7}
\plotone{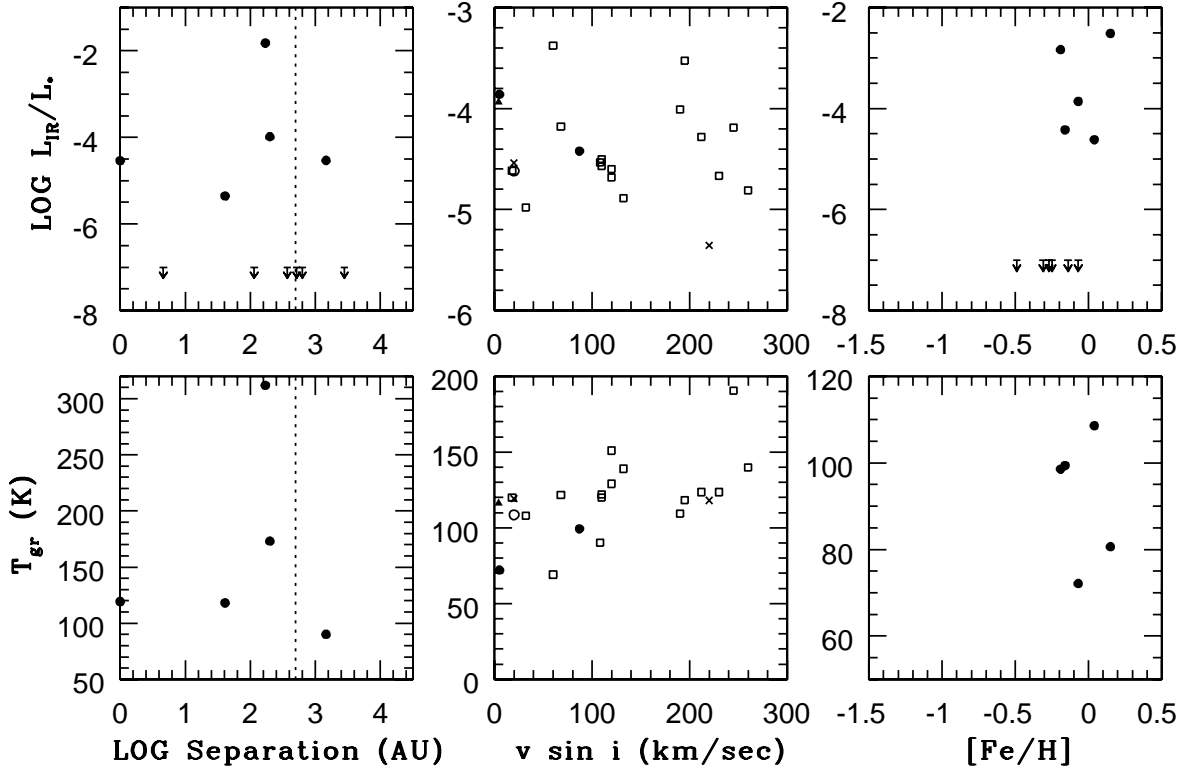}
\caption{Inferred fractional luminosity, $L_{IR}/L_{*}$, and grain temperature,
$T_{gr}$, plotted as a function of binary separation (for all binaries in our 
study), measured stellar rotational velocity, $v \sin i$, and stellar 
metallicity, [Fe/H]. For binary systems to the left of the dotted line, the LL 
(and sometimes the SL) slit contain both the primary and secondary; for
objects to the right of the dotted vertical line, the SL and LL slits contain 
only the primary star. In plots of grain properties as a function of 
$v \sin i$ and [Fe/H], B-type stars are shown as crosses, A-type stars are
shown as squares, F-type stars are shown as circles, and K-type stars are
shown as triangles. Stars with spectral type earlier than F5V are shown as
open symbols in $v \sin i$ plots. Stars with spectral type earlier than
F0V are shown with open symbols in [Fe/H] plots.}
\end{figure}

\end{document}